\documentclass[fleqn,usenatbib]{mnras}

\usepackage{newtxtext,newtxmath}

\usepackage[T1]{fontenc}

\DeclareRobustCommand{\VAN}[3]{#2}
\let\VANthebibliography\thebibliography
\def\thebibliography{\DeclareRobustCommand{\VAN}[3]{##3}\VANthebibliography}

\usepackage{graphicx}	
\usepackage{amsmath}	
\usepackage{amssymb}	

\newcommand{\wj}[6]{ \begin{pmatrix} 
   #1 & #2 & #3 \\
   #4 & #5 & #6 
  \end{pmatrix}}

\title[Galaxy window function]{On the impact of the galaxy window function on cosmological parameter estimation}

\author[T. Karim et al.]{
Tanveer Karim,$^{1}$\thanks{E-mail: tanveer.karim@cfa.harvard.edu (TK)}
Mehdi Rezaie,$^{2, 3}$
Sukhdeep Singh$^{4}$,
Daniel Eisenstein$^{1}$
\\
$^{1}$Center for Astrophysics | Harvard \& Smithsonian, 60 Garden St, MS 10, Cambridge, MA 02138, USA\\
$^{2}$Department of Physics, Kansas State University, 116 Cardwell Hall, Manhattan, KS 66506, USA\\
$^{3}$Center for Cosmology and AstroParticle Physics, The Ohio State University, 191 West Woodruff Avenue, Columbus, OH 43210, USA\\
$^{4}$McWilliams Center for Cosmology, Department of Physics, Carnegie Mellon University, Pittsburgh, PA 15213, USA
}

\date{Accepted 2023 July 18. Received 2023 July 12; in original form 2023 May 19}

\pubyear{2023}

\begin{document}
\label{firstpage}
\pagerange{\pageref{firstpage}--\pageref{lastpage}}
\maketitle

\begin{abstract}
One important source of systematics in galaxy redshift surveys comes from the estimation of the galaxy window function. Up until now, the impact of the uncertainty in estimating the galaxy window function on parameter inference has not been properly studied. In this paper, we show that the uncertainty and the bias in estimating the galaxy window function will be salient for ongoing and next-generation galaxy surveys using a simulation-based approach. With a specific case study of cross-correlating Emission-line galaxies from the DESI Legacy Imaging Surveys and the \emph{Planck} CMB lensing map, we show that neural network-based regression approaches to modelling the window function are superior in comparison to linear regression-based models. We additionally show that the definition of the galaxy overdensity estimator can impact the overall signal-to-noise of observed power spectra. Finally, we show that the additive biases coming from the window functions can significantly bias the modes of the inferred parameters and also degrade their precision. Thus, a careful understanding of the window functions will be essential to conduct cosmological experiments.
\end{abstract}




\section{Introduction}
\label{sec3:intro}

The matter density field encodes critical cosmological information pertaining to cosmological properties and the dark sector. As observers, we are interested in measuring statistics of this field as a function of redshift. Galaxy redshift surveys use galaxies as tracers of the matter density field to measure these statistics. The galaxy redshift surveys count the number of galaxies in different patches of the sky and measure the clustering of the galaxy overdensity field, which can then be translated to the underlying matter density field assuming models of galaxy-dark matter connection, known as the galaxy bias. Thus, from an observer's point of view, measuring the true galaxy clustering is one of the few observables that allows us to probe cosmological models. But observational effects, such as Galactic extinction, seeing conditions, telescope response at various wavelengths, and the survey geometry, can modulate the true galaxy number count. Hence, if we do not understand how these observational effects impact the true galaxy number count, it can lead to a biased measurement of the underlying matter density field, and by extension lead to the incorrect measurements of cosmological parameters \citep{Huterer2013, Thomas2011}. 

The galaxy window function is a way to encode such observational systematics and the survey geometry; it can be thought of as a weighting function that tells us how the true galaxy number count is being modulated as a function of right ascension and declination (and redshift if measuring $3$D clustering) on the sky. Since the window function depends entirely on observational effects, the detector and the survey geometry, we cannot determine the window function from first principles. Consequently, it must be modelled carefully using templates to null out its effects. While the conventional approach is to make a point estimate of the window function by regressing over the templates \citep{Elvin-Poole18, Ross12}, such an approach fails to take into account the inherent uncertainty in estimating the window function and its variance. Especially as we enter the era of sub-per cent level precision cosmology, it has become critical to assess whether the variance and uncertainty in estimating the galaxy window function can be one of the leading sources of systematic errors; such issues are beginning to be studied as the precision of the surveys increases over time \citep{weaverdyck21, Singh21}. 

This paper aims to address this question with the help of Gaussian mocks of the galaxy overdensity field and the galaxy window function. We specifically model how additive and multiplicative biases in estimating the window function can impact cosmological parameters such as the amplitude of the matter power spectrum, $A_s$, and the galaxy linear bias, $b_0$. We investigate this problem with a case study of the cross-correlation of Emission-line galaxies (ELGs) from the DESI Legacy Imaging Surveys and the Planck CMB lensing. This question is especially timely because in recent years, a persistent mild tension in the amplitude of the matter power spectrum, known as the $\sigma_8$ tension, has appeared in the literature \citep{valentino2021s8, Nunes21}. Specifically, the measurements of the amplitude based on CMB data \citep{planck2020, aiola2020} is at $2 - 3\sigma$ tension with low-redshift probes such as weak lensing, cosmic shear, cluster counting, redshift-space distortions (RSD), full-shape power spectrum and cross-correlation of large-scale structure with CMB lensing (For a detailed discussion, refer to Section V of \citet{abdalla2022}). We seek to address this important problem by cross-correlating the largest catalogue of emission-line galaxies (ELGs), the DESI Legacy Imaging Surveys, and the \emph{Planck} CMB lensing map in an in-preparation companion paper \citep{Karim2023}. While the mild tension could point towards the evidence of "new physics", it is also possible that the tension can be explained away by accounting for previously unaccounted systematics. Thus, this paper provides a pathway for our companion paper, as well as future cosmological survey papers, to consider a new source of systematic uncertainties. While this paper specifically focuses on the angular clustering of galaxies and galaxy-CMB lensing, these ideas can be extended to full $3$D clustering and will be explored in a future paper with DESI spectroscopic data.

We specifically try to answer the following three questions in this paper: 

\begin{itemize}
    \item How much does the bias and the variance of the window function impact parameter inference?
    \item Are linear regression-based approaches to estimating the window function sufficient?
    \item How much (if any) does the definition of the galaxy overdensity field estimator affect inference? 
\end{itemize}

To answer these questions, the paper is organized as follows. In Section~\ref{sec3:winfunc}, we discuss the definition and mathematical formalism of the window function. We also discuss how the definition of the window function can affect the galaxy overdensity map estimator, $\hat{\delta}_g$. Section~\ref{sec:data} describes the data we used for our analysis, and Section~\ref{sec:selfunc} discusses two different approaches to modelling the galaxy window function, namely the linear regression method and the neural network method. In Section~\ref{sec:method} we describe the procedure to generate the mocks used in this analysis. Section~\ref{sec:result} discusses our key findings, specifically why a neural network-based approach to modelling the galaxy window function is superior to a linear regression approach, as well as comparisons of how the different definitions of the window function affect the power spectra measurement and parameter inference. Finally, in Section~\ref{sec:conc} we provide a summary of this paper and contextualise our findings in the context of ongoing and future surveys.

\section{Window Function}
\label{sec3:winfunc}

Angular clustering of galaxies can be measured either by counting pairs or by discretising the sky into pixels and then measuring the clustering of the effective galaxy overdensity field values coming from each of those pixels. In this paper, we focus on the latter approach and use the commonly used pixelization scheme, \textsc{healpix} \citep{Gorski05, Zonca2019}. This approach effectively treats each pixel as having the same weight when measuring the power spectra. However, foreground artefacts, e.g., Galactic extinction, may affect the survey footprint non-uniformly and lead to a systematic undercounting of the overdensity field. For example, a faint galaxy may not get observed if it is in a pixel closer to the Galactic Plane due to extinction. Hence, we need a mechanism to quantify the weights of a given survey pixel. 

The galaxy window function (hereafter window function or $W_g$) is a weighting function that quantifies the mean number of galaxies expected to be observed if the Universe was unclustered, by regressing over systematics template maps. It captures and encodes two different effects --  the survey geometry (or the mask) and the observational effects present within the survey footprint. Note that while our discussion in this paper is limited to angular clustering, it can be extended to full $3$D treatment for spectroscopic or tomographic surveys. 

Because of the observational effects and survey geometry, the measured angular clustering from the sky is a convolution of the underlying clustering signal coming from Large-Scale Structures and the window function. Thus, when it comes to cosmological analysis, one must model the window function carefully to deproject its contribution from the measured angular clustering (or forward model its contribution). 

Typically the actual observable that an observer measure is the number of galaxies per pixel. The observer then has to convert the raw number count to a galaxy overdensity field which is used to measure clustering. In Section~\ref{sec3:mathform} we discuss how the overdensity field, $\delta_g$, is estimated in the presence of a window function.

\subsection{Mathematical Formalism}
\label{sec3:mathform}
  The galaxy overdensity field, ${\delta}_g$, is estimated using the discretized number counts per pixel that are used in the actual power spectrum (or higher-order statistics) analysis. However, as the actual observation is affected by the window function, the estimator needs to account for the window function either implicitly or explicitly. The former approach has been used extensively in past surveys \citep{Ross12, Elvin-Poole18}. 

 In the explicit approach, the windowed estimator of the galaxy overdensity field, $\hat{\delta}_{g,W}$, accounts for both the window function, $W_g$, and the true galaxy overdensity field, ${\delta}_g$. Following the convention of \citet{Singh21}, let us imagine a galaxy survey where we count, $n_g (x)$, the number of galaxies per pixel, and $x$ denotes the angular position of the pixel on the sky. We can express this quantity as:

 \begin{equation}
 \label{eq3:ng}
     n_g (x) = \left< n_g (x) \right> \left[1 + \delta_g (x) \right]
 \end{equation}
 
\noindent where $\left< n_g (x) \right>$ is the ensemble average of $n_g (x)$, i.e., the expected number of galaxies in absence of clustering and noise. Now, if we measure the average number of galaxies per pixel, $\bar{n}_g$, in our survey, then we can define the windowed estimator of the galaxy overdensity field as:

\begin{align}
    \hat{\delta}_{g,W} &= \frac{n_g (x)}{\bar{n}_g} - \frac{\left <n_g (x) \right>}{\bar{n}_g} \label{eq3:explicit_alt} \\
    &= \frac{\left <n_g (x) \right> \left[1 + \delta_g (x) \right]}{\bar{n}_g} - \frac{\left <n_g (x) \right>}{\bar{n}_g} \\ 
    &= \frac{\left <n_g (x) \right>}{\bar{n}_g} \delta_g (x) \\
    &= W_g (x) \delta_g (x) \label{eq3:explicit}
\end{align}

 \noindent where in the last equality we define the galaxy window function as $W_g (x) = \frac{\left <n_g (x) \right>}{\bar{n}_g}$. Thus, another way to think about the galaxy window function is that it quantifies the modulation of the ensemble average not due to clustering or noise, but due to other sources. 



In contrast, in the implicit approach, the observed overdensity map is divided by the estimated window function to remove its effects. This can be thought of as dividing the windowed estimator from Equation~\ref{eq3:explicit} by the window function, such that:

\begin{equation}
\label{eq3:implicit}
    \hat{\delta}_{g} (x) = \frac{\hat{\delta}_{g,W} (x)}{W_g (x)} =  \frac{W_g (x) \delta_g (x)}{W_g (x)} =  \delta_g (x)
\end{equation}

However, in this approach, the mask (or the survey geometry) still needs to be accounted for and the final mask needs to be more conservative to remove pixels where the window function is small. 

Note that in a real survey, we also have to account for the galaxy shot noise. For an all-sky survey with an average number of galaxies per pixel, $\bar{n}_g$, the shot noise power spectrum is given by, 

\begin{equation}
\label{eq3:ngg}
    N^{gg}_{\ell} = \frac{1}{\bar{n}_g}
\end{equation}

However, in the presence of galaxy windows and masks, the contributions of the shot-noise in $\hat{\delta}_g (x)$ for the implicit and explicit approaches are given as \citep{Singh21}:

\begin{align}
    \langle\delta_N^2\rangle(x) &= \overline{\left[ \frac{1}{W_g (x)} \right]} N^{gg}_{\ell} \label{eq:implicit_noise} \\
    \langle\delta_{N,W}^2\rangle(x) &= \overline{W}_g (x)  N^{gg}_{\ell} \label{eq:explicit_noise}
\end{align}

\noindent where, $\langle\delta_N^2\rangle(x)$ denotes the ensemble average of the square of the noise map realizations, and $\overline{W}_g (x) = 1$ by construction. Notice that  Equations~\ref{eq:implicit_noise} and \ref{eq:explicit_noise} refer to the ensemble average of the square of the galaxy window function noise. Thus, to obtain the noise estimates, we have to take the square root of the right-hand sides. Furthermore, the left-hand side refers to the ensemble averages. However, when we consider realizations, the noise power spectra on the right-hand side refer to realizations of all-sky noise maps whose power spectra are Equation~\ref{eq3:ngg}. 

Therefore, if an all-sky realization of the shot-noise power spectra, $N^{gg}_{\ell}$ is given by $\delta^{N}_{g} (x)$, the implicit and the explicit approaches are in reality measuring:

\begin{align}
    \hat{\delta}_g (x) &= \delta_g (x) + \delta_g^N (x) \sqrt{\overline{\left[ \frac{1}{W_g (x)} \right]}} \label{eq:implicit}\\ 
    \hat{\delta}_{g,W} (x) &= W_g (x) \delta_g (x) + \delta_g^N (x) \sqrt{\overline{W}_g (x)}  \label{eq:explicit}
\end{align}

One should note that due to Jensen's inequality, $\overline{\left[ \frac{1}{W_g (x)} \right]} \geq \overline{W}_g (x)$, the explicit estimator is less noisy than the implicit estimator in the shot-noise dominated regime.

\subsection{Pseudo-$C_{\ell}$ framework}
\label{sec:pcl}

The angular power spectra, $C_{\ell}$, can be calculated analytically from a cosmological model if the redshift kernel of the tracer is known \citep{Limber1953}. From the observation side, the same power spectra can be measured from an all-sky survey. Thus, if one could observe the full sky, then inference on cosmological parameters would be trivial.

However, the presence of the window function and the survey mask complicates this analysis. The window function induces mode coupling in the observed power spectra, which is also known as the pseudo-$C_{\ell}$ or $D_{\ell}$. While $C_{\ell}$ measures the variance of the underlying overdensity field, $D_{\ell}$ measures the variance of the estimators described in Equations~\ref{eq:implicit} and \ref{eq:explicit}, depending how one assigns weights to the map. The relationship between $C_{\ell}$ and the estimator of $D_{\ell}$ are given by \citep{Hivon02}:

\begin{equation}
\widehat{D}_{\ell} = M_{\ell\ell'}C_{\ell'} \label{eq:pcl}
\end{equation}

\noindent where, $M_{\ell\ell'}$ is the mode coupling matrix. The coupling matrix is given by, 

\begin{align}
 	M_{\ell,\ell'}={\frac{(2\ell'+1)}{4\pi}}\sum_{\ell''}
     W_{\ell''}(2\ell''+1)
    &\wj{\ell}{\ell'}{\ell''}{s_1}{-s_1}{0}\nonumber \\ \times&\wj{\ell}{\ell'}{\ell''}{s_2}{-s_2}{0}
\label{eq:couplingM}
\end{align}
\noindent where $W_{\ell''}$ is the angular power spectra of the window function, $s_1,s_2$ are the spins of the two tracers being correlated to measure $\widehat{D}_{\ell}$, and the expression in the parenthesis on the right are the Wigner 3-j symbol \citep{Wigner93}. For galaxies and CMB lensing, the tracer spin values are $0$. This effectively shows that power from small scales can leak into large scales and mischaracterization of the window function can lead to a biased estimate of the underlying cosmology, even if the analysis is not focused on the largest scales. Note that here we assume that the survey geometry or mask and the tracer fields are uncorrelated. If there is a correlation, then a modification of the coupling matrix equation can be modified appropriately \citep{surrao2023}. Even in such cases, the arguments put forward in this paper can be similarly used. 

While in theory, one can estimate $C_{\ell}$ by inverting Equation~\ref{eq:pcl}, in practice, this leads to a lossy estimate, especially in the presence of a noise power spectra with a large amount of power at high multipoles, e.g., the CMB lensing noise power spectra. Thus, forward modelling the $D_{\ell}$ from the underlying cosmology can be a more robust way of measuring cosmological parameters. As a result, in the rest of the paper we measure the observed $\widehat{D}_{\ell}$ from mocks and compare it against forward modelled $D^{\rm th}_{\ell}$ to do our inference analysis.

\subsection{Impact of Bias and Variance of the Window Function}
\label{ssec:window_bias_model}

As discussed in Section~\ref{sec3:mathform}, there are two ways of accounting for the galaxy window function when estimating the galaxy overdensity field. We can use either of these approaches to measure the galaxy overdensity field if we know the exact form of the galaxy window function. But in reality, the underlying true galaxy window function is never known, but estimated. This raises an important question -- \textit{how do the bias and variance of the window function estimator affect the galaxy overdensity estimator, and ultimately the cosmological parameters of interest?} 

We investigate this question by explicitly modelling the additive and multiplicative biases of the window function, as well as the variance of the window using a simulation-based approach (Section~\ref{sec:mocks}). Modelling of the additive and multiplicative biases is important because even if we understand the foreground artefacts or imaging systematics exactly, the estimation of the window function from the foreground artefact maps is a non-trivial task and can result in some errors. This error could have both an additive component, i.e., the estimated window function is shifted from the true window function by a certain value per pixel, and it could also have a multiplicative component, i.e., the estimated window function is scaled by a certain value per pixel with respect to the true window function. 

To further motivate the physical interpretation of the difference between the additive and multiplicative biases, let us assume that the galaxy window function estimator, $\hat{W}_g$, can be represented as a function of the true galaxy window function, $W_{g}$ in the following way:

\begin{equation}
\label{eq3:biased_wg}
    \widehat{W}_g = \left[1 + m(x) \right] W_{g} (x)
\end{equation}

\noindent where $m(x)$ is the relative error in estimating the galaxy window function. Due to the error in the galaxy window function estimator, the corresponding windowed galaxy overdensity field estimator will also be biased. Using Equation~\ref{eq3:explicit_alt}, we can write down the biased windowed galaxy overdensity field estimator, $\hat{\delta}_{g,W,m}$, as \citep{Singh21}:

\begin{align}
    \hat{\delta}_{g,W,m} = \frac{n_g(x)}{\bar{n}_g} - \frac{\left< \widehat{n_g (x)} \right>}{\bar{n}_g} \label{eq3:biased}
\end{align}

\noindent where the second term is referring to the biased galaxy window function estimator $\widehat{W}_{g}$. We can rewrite Equation~\ref{eq3:biased} by using Equations~\ref{eq3:ng} and ~\ref{eq3:biased_wg}:

\begin{align*}
    \hat{\delta}_{g,W,m} &= \frac{\left< n_g (x) \right> \left[1 + \delta_g (x) \right]}{\bar{n}_g} - \widehat{W}_g \\ 
    &= W_g (x) \left[1 + \delta_g (x) \right] - \widehat{W}_g \\
    &= W_g (x) + W_g (x) \delta_g (x) - W_g (x) - m(x) W_g (x) \\
    &= W_g (x) \delta_g (x) - m(x) W_g (x) \\
    &= \frac{\widehat{W}_g}{1 + m(x)} \delta_g (x) - m(x) W_g (x) \\
    &= \widehat{W}_g \delta_g - \widehat{W}_g m(x) \delta_g (x) - m(x) W_g (x)
\end{align*}

\noindent where the last equality holds by Taylor series expansion of $1/(1 + m(x))$ around $m(x) = 0$. Thus, we see that while the multiplicative bias scales the overdensity field by a factor of $\widehat{W}_g m(x)$, the additive bias term (the last term) is independent of $\delta_g (x)$. 

For the implicit approach, one can also similarly show that:

\begin{align*}
    \hat{\delta}_{g, m} (x) = \frac{\delta_g (x) - m(x)}{1 + m(x)}
\end{align*}

With the multiplicative and additive biases defined, we can now model the impact of the additive and multiplicative biases on both the explicit and implicit approaches.

Let us assume that $W_{g,t}$ is the true window function and $W_{g,e}$ is an estimation of the window function based on foreground systematics template maps. For the explicit approach, we can then model the windowed true, multiplicative-biased, and additive- and multiplicative-biased galaxy overdensity estimators as:

\begin{align}
    \hat{\delta}^{A}_{g,W} (x) &= W_{g,t} (x) \delta_g (x) + \delta^{N}_g (x) \sqrt{\overline{W}_{g,t} (x)} \label{eq3:expA} \\
    \hat{\delta}^{B}_{g,W} (x) &= W_{g,e} (x) \delta_g (x) + \delta^{N}_g (x) \sqrt{\overline{W}_{g,e} (x)} \label{eq3:expB} \\
    \hat{\delta}^{C}_{g,W} (x) &= W_{g,e} (x) \left(1 + \delta_g (x) \right) + \delta^{N}_g (x) \sqrt{\overline{W}_{g,e} (x)} - W_{g,t} (x) \label{eq3:expC}
\end{align}


\noindent where A, B, and C are the labels for these models. As $W_{g,e}$ approaches $W_{g,t}$, Equations~\ref{eq3:expB} and \ref{eq3:expC} approach Equation~\ref{eq3:expA}. In the case of Equation~\ref{eq3:expB}, the modulation of $\hat{\delta}^{B}_{g,W} (x)$ with respect to $\hat{\delta}^{A}_{g,W} (x)$ denotes what is the multiplicative bias effect because correcting for the multiplicative bias effectively means rescaling the right-hand side of Equation~\ref{eq3:expB}. On the other hand, Equation~\ref{eq3:expC} contains both the right-hand side from Equation~\ref{eq3:expB} and an additional term of $W_{g,e} (x) - W_{g,t} (x)$. This extra term is the is the additive component in Equation~\ref{eq3:expC}. If $W_{g,e} (x)$ were exactly equal to $W_{g,t} (x)$, then this term would vanish and we would get back the usual Equation~\ref{eq3:expA}.

Following a similar reasoning, for the implicit approach, we model the true, multiplicative-biased, and additive- and multiplicative-biased galaxy overdensity estimators as:

\begin{align}
    \hat{\delta}^D_g (x) &= \delta_g (x) + \delta^{N}_g (x) \sqrt{\overline{\left[ \frac{1}{W_{g,t} (x)} \right]}} \label{eq3:expD} \\
    \hat{\delta}^{E}_{g} (x) &= \frac{W_{g,e} (x)}{W_{g,t} (x)} \delta_g (x) + \delta^{N}_g (x) \frac{\sqrt{\overline{W}_{g,e} (x)  }}{W_{g,t} (x)} \label{eq3:expE} \\
       \hat{\delta}^{F}_{g} (x) &= \frac{W_{g,e} (x)}{W_{g,t} (x)} \left( 1 + \delta_g (x) \right) + \delta^{N}_g (x)\frac{\sqrt{\overline{W}_{g,e} (x)}}{W_{g,t} (x)} - 1 \label{eq3:expF}
\end{align}

\noindent where D, E, and F are the labels for these models. These equations are obtained by dividing Equations~\ref{eq3:expA} -- \ref{eq3:expC} by $W_{g,t} (x)$ as in the implicit approach we divide out the effect of the galaxy window function from the estimator. Again, we see that as $W_{g,e}$ approaches $W_{g,t}$, Equations~\ref{eq3:expE} and \ref{eq3:expF} approach Equation~\ref{eq3:expD}. Note that in reality, an observer can only measure either Equation~\ref{eq3:expC} or \ref{eq3:expF}. 

But, we also include variants expressed in Equations~\ref{eq3:expA} and \ref{eq3:expD} to showcase what our best performance could be if there was a way to know the true galaxy window function. These variants define the benchmark for the best performance and allow us to compare the impact of additive and multiplicative biases. Similarly, we include variants expressed in Equations~\ref{eq3:expB} and \ref{eq3:expE} to understand how much of an impact multiplicative biases have on the estimators, if we knew the amount of additive bias \emph{a priori}. Comparing Equations~\ref{eq3:expB} to \ref{eq3:expC} or Equations~\ref{eq3:expE} to \ref{eq3:expF} enables us to isolate the contributions of multiplicative biases from additive biases.    

Thus, if we generate Gaussian realization of the same underlying cosmology, and apply these various definitions, we can then use our simulation-based approach to explicitly quantify how much these biases impact cosmological inference as well as answer which of the two approaches increases the overall signal-to-noise. Additionally, we can also investigate two popular choices of the galaxy window function estimator (detailed discussion in Section~\ref{sec:selfunc} -- linear regression and neural network, to assess which method reduces the overall bias and variance the most. 

\section{Data}
\label{sec:data}

We use two different surveys and their cross-correlation measurements to understand the impact of the galaxy window function estimation. These surveys are described in detail below. 
 
\subsection{DESI Legacy Imaging Surveys DR9 Catalogue}
The DESI Legacy Imaging Surveys is a combination of three photometric surveys -- the Dark Energy Camera Legacy Survey (DECaLS), the Beijing-Arizona Sky Survey (BASS) and the Mayall $z$-band Legacy Survey (MzLS). Together, they cover $\sim 14000$ deg$^2$ of the northern hemisphere sky in $g$ $r$ and $z$ optical bands and four additional infrared bands \citep{Dey19}. The DESI Legacy Imaging Surveys serves as the input catalogue for the DESI target selection algorithms; as all DESI targets are pre-selected, the DESI target selection algorithms require good photometry to determine which objects will meet the DESI science goals. Thus, all the DESI emission-line galaxies targets form a subset of all the emission-line galaxies observed in the DESI Legacy Imaging Surveys.  

Because the DESI spectroscopic survey relies on the DESI Legacy Imaging Surveys for target selection, it inherits all the associated imaging systematics and target density systematics from the Legacy Surveys. Hence, understanding imaging systematics and the galaxy window function is important not only for any DESI Legacy Imaging Surveys specific science analysis but also critical for any future DESI analysis. In total, the DESI Legacy Imaging Surveys Data Release 9 (DR9) catalogue contains $\sim 22$ million ELGs (after applying colour and magnitude cuts) spanning $\sim 40\%$ of the sky.  

For this paper, we use the DR9 dataset\footnote{https://www.legacysurvey.org/dr9/}. The colour selection of the ELGs is described in detail in \citet{Karim2023}. 

\subsection{\emph{Planck} CMB Lensing Map}
To generate \emph{Planck} CMB-like mocks, we use the mask and the noise model of the 2018 data release\footnote{http://pla.esac.esa.int/pla/} \citep{planck2018}. The CMB lensing map traces the distortion of the CMB photons along the line of sight as they encounter the gravitational potential of masses. Depending on the geometry between the surface of the last scattering and the observer, the gravitational potential can magnify or suppress the CMB photons, distorting the observed temperature and polarization measurements. We use the noise model from the SMICA DX12 CMB minimum-variance estimate maps and obtain it from the \textsc{com\_lensing\_4096\_r3.00} dataset. This dataset provides the noise model up to $\ell_{\rm max} = 4096$. Note that \emph{Planck} provides their data in the Galactic Coordinate System, while we perform our analysis in the Equatorial Coordinate System. Thus, we rotate the \emph{Planck} mask from the Galactic Coordinate System to Equatorial Coordinate System before using it. 

\section{Modelling Galaxy Window Function}
\label{sec:selfunc}

The Galaxy window function is derived by modelling the observed number count of galaxies in pixel \textit{i}, $n_{g, i}$, given a set of imaging systematics maps, $\textbf{s}_{i}$. The underlying cosmological signal is assumed to not correlate with foreground imaging maps, and thus the regression analysis returns the window function that encapsulates large-scale spurious fluctuations caused by varying imaging foreground systematics. As this is essentially a regression problem, we investigate both the linear regression and the neural network method. While the linear regression method makes the result interpretable, the neural network approach can map any non-linearities that may exist between the background large-scale structure and the foreground imaging systematics. 

In our analysis, maps of imaging properties are produced in \textsc{healpix} with \textsc{nside}=$1024$ from the catalogue of random galaxies with similar angular masking, and include \textit{galactic extinction} \citep{Schlegel98}, stellar density \citep{gaia18}, and the galaxy depth, point spread function (PSF) depth, and the PSF size in three optical bands, i.e., \textit{r}, \textit{g}, and \textit{z}. Thus, we have a total of $11$ foreground systematics maps. 

\begin{figure}
    \centering
    \includegraphics[page=1, width=0.4\textwidth]{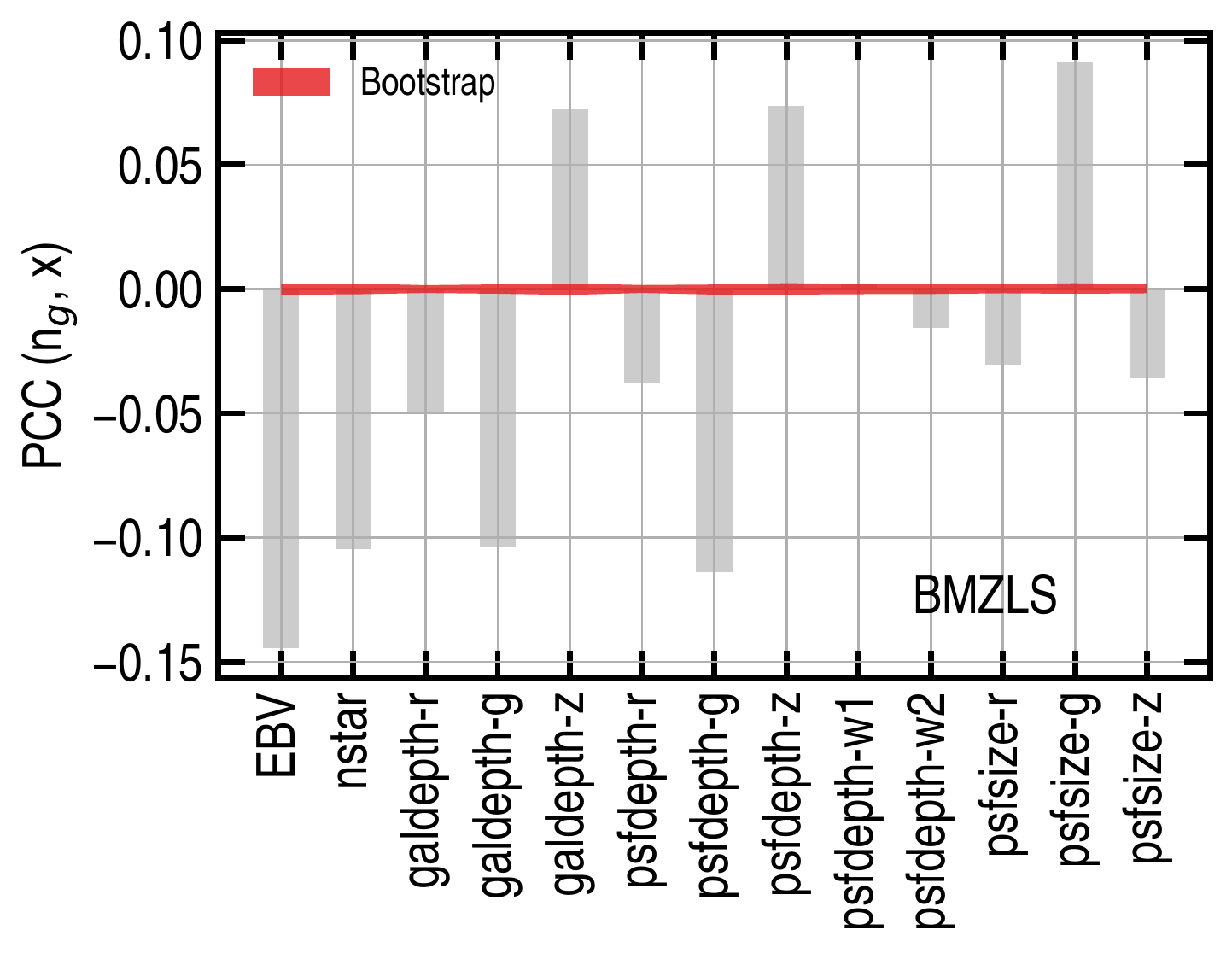}
    \includegraphics[page=2, width=0.4\textwidth]{figures/pcc_dr9elg.pdf}
    \includegraphics[page=3, width=0.4\textwidth]{figures/pcc_dr9elg.pdf}
    \caption{Pearson's correlation coefficients between the DESI ELG density field and imaging properties in the BASS/MzLS, DECaLS North, and DECaLS South footprints, respectively from top to bottom. The horizontal shade illustrates the range of variations in 100 Bootstrapped realizations.}
    \label{fig:pcc_dr9elg}
\end{figure}

We find that although the DESI Legacy Imaging Surveys pipeline tried to uniformize the photometry of the distinct regions (BASS/MzLS, Northern DECaLS and Southern DECaLS) as much as possible, the difference between the regions is still apparent. Figure~\ref{fig:pcc_dr9elg} shows the Pearson correlation coefficient (PCC) between the ELG number density and the foreground systematics maps in the three regions; the horizontal shades are $100$ bootstrapped realizations. If there were no foreground systematics effects, then the PCC would have been zero everywhere. However, the differences in the correlation in the three regions indicate that the window functions have to be modelled separately in the three regions first, before combining them together. We combine them by normalizing their predicted mean density with a global mean.

\subsection{Regression of ELG Density}
The modelling is performed with a neural network (see, \ref{ssec:nn})  and a linear multivariate model (see, \ref{ssec:lin}) as the baseline approach for benchmarking. The parameters of each model are trained by minimizing the negative Poisson log Likelihood as \textit{loss} function $J$ \citep[see, e.g.,][]{rezaie2021MNRAS.506.3439R},
\begin{equation}\label{eq:cost}
    J = \sum_{i}{f_{{\rm pix}, i}[f_{{\rm pix},i} \mathcal{Y}_{i}-n_{g, i}\log(f_{{\rm pix},i} \mathcal{Y}_{i})]}
\end{equation}
where $\mathcal{Y}_{i}\equiv \mathcal{Y}(\textbf{s}_{i}|\Theta)$ represents the modelled galaxy window function in pixel $i$ and $f_{{\rm pix},i}$ describes survey completeness in pixel $i$, regardless of imaging effects, and is determined from projecting a catalogue of random galaxies onto \textsc{healpix}.

\subsubsection{Linear Regression based approach}\label{ssec:lin}
\begin{figure}
    \centering
    \includegraphics[width=\columnwidth]{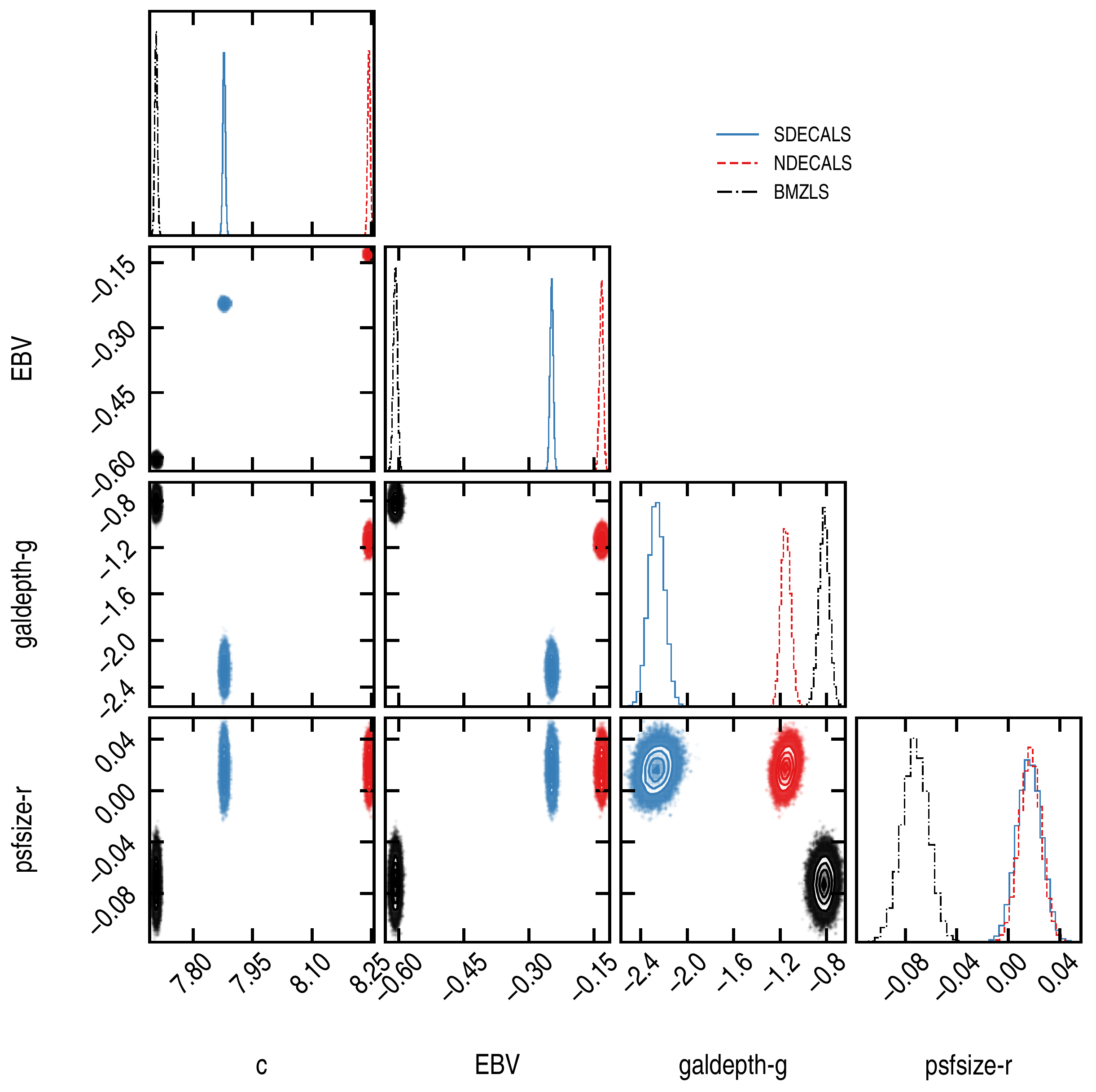}
    \caption{Posterior of linear model parameters based on Eq. \ref{eq:linmodel}. The plots show that the three regions of the DESI Legacy Imaging Surveys have vastly different foreground imaging systematics.}
    \label{fig:linear_reg_mcmc}
\end{figure}

Our first approach uses linear regression to model the galaxy window function. We use the likelihood defined in Eq. \ref{eq:cost} with the Markov Chain Monte Carlo (MCMC) sampler, \textsc{emcee} \citep{emcee}, to probe the parameter space of a linear model,
\begin{equation}\label{eq:linmodel}
    \mathcal{Y}(\textbf{s}_{i}|\Theta) = \log \left[ 1 + \exp{\left[ c + \sum_{j} \theta_{j} \left( \frac{s_{i, j}-\mu_{j}}{\sigma_{j}} \right) \right]} \right],
\end{equation}
where $c$ is the intercept, $s_{i, j}$ is the $j$'th imaging map in pixel $i$, and $\mu_{j}$ and $\sigma_{j}$ are respectively its estimated \textit{mean} and \textit{standard deviation}. We use flat priors for all parameters and initialise 400 MCMC chains for 1000 steps, 400 of which are left aside as the burn-in phase. We then randomly sample 1000 points in the parameter space to generate the ensemble of linear window functions. The posterior of some of the linear model parameters are shown in Figure~\ref{fig:linear_reg_mcmc}. Note that Equation~\ref{eq:linmodel} is technically a soft-plus function to ensure that the output is always positive, and that this function is better suited for Poisson processes. If the argument in the exponential is large, then this function reduces to a linear model (or to zero if the argument is small). Thus, we use the term "linear model" to refer to this soft-plus function. 

\subsubsection{Neural Network based approach}\label{ssec:nn}
Next, we use a neural network-based regressor to model the galaxy window function. The rationale behind using a neural network regressor is that, unlike a linear model, a neural network model can capture non-linear mapping between the foreground systematics maps and the galaxy window function. 

The architecture of our feed-forward neural network has three hidden layers and $20$ neurons on each layer. The input layer takes $11$ imaging properties and the output layer is composed of a single neuron which returns galaxy window function $\mathcal{Y}$. In the feed-forward architecture, the value of neuron $m$ in layer $l$, $a_{m}^{l}$, is related to the values from layer $l-1$ via,
\begin{equation}\label{eq:fnn}
    a_{m}^{l} = f(c^{l}_{m}+\sum_{j}\theta^{l}_{m, j}a_{j}^{l-1})
\end{equation}
where $c^{l}$ and $\theta^{l}$ are the bias and weight parameters associated to layer $l$, and $f$ is the activation function used in the neuron. Specifically, we have $f(u) = max(0, u)$ in the hidden layer neurons and $f(u)=\log(1+\exp(u))$ in the last layer neuron. Imaging properties are propagated to the galaxy window function via the recursive relation in Eq. \ref{eq:fnn}. The parameters $\Theta$, which includes biases and weights for all layers, are trained iteratively \citep{Loshchilov2017arXiv171105101L},
\begin{align}
    \Theta_{t+1} &= \Theta_{t} - \eta_{t} \frac{m_{t+1}}{\sqrt{\nu_{t+1}}+\epsilon},
\end{align}
where $\epsilon=10^{-8}$. The first and second moments of the cost function gradient, $m_{t}$ and $\nu_{t}$, are both initialised as zero and changed iteratively via, 
\begin{align}
    m_{t+1} &= \beta_{1}m_{t}+(1-\beta_{1})\nabla J, \\
    \nu_{t+1} &= \beta_{2}\nu_{t} + (1-\beta_{2})|\nabla J|^{2},
\end{align}
with parameters $\beta_{1}=0.9$ and $\beta_{2}=0.999$. The learning rate parameter $\eta_{t}$ is designed to alternate between $\eta_{\rm min}$ and $\eta_{\rm max}$ every five epochs \citep{Loshchilov2016arXiv160803983L},
\begin{equation}
\eta_{t} = \eta_{\rm min} + \frac{1}{2}(\eta_{\rm max} - \eta_{\rm min}) [1 + \cos(\frac{t_{5}}{5}\pi)],
\end{equation}
where $t_{5}$ is the number of epochs since the last restart, and the optimal range for learning rate, $\eta_{\rm min}=10^{-5}$ and $\eta_{\rm max}=0.01$, is determined via a grid search \citep{smith2015arXiv150601186S}. We train the neural network with $60\%$ of the data while leaving $20\%$ for validation and $20\%$ for testing. The training, validation and testing set split is performed randomly to ensure that the distributions of the imaging properties in the training set match those in the validation and test sets. This kind of split ensures that the model does not have to extrapolate in the imaging space when it is applied to the validation and test sets. Additionally, using a contiguous split would introduce large-scale structure clustering when we regress on the on-sky real data. This would violate the assumption of independence of the training pixels from each other and as a result, our cost function would no longer be valid. Thus, the splitting on the basis of random pixels is the best choice for our problem.

We train the network for 200 epochs while saving a snapshot of the network every five epochs. This technique allows us to create an ensemble of 40 neural networks with one training run \citep[see, e.g.,][]{huang2017arXiv170400109H}. We increase the ensemble size to 1000 by re-partitioning the training-validation-testing splits five times and initialising five different networks for each split. 

\begin{figure}
    \centering
    \includegraphics[width=\columnwidth]{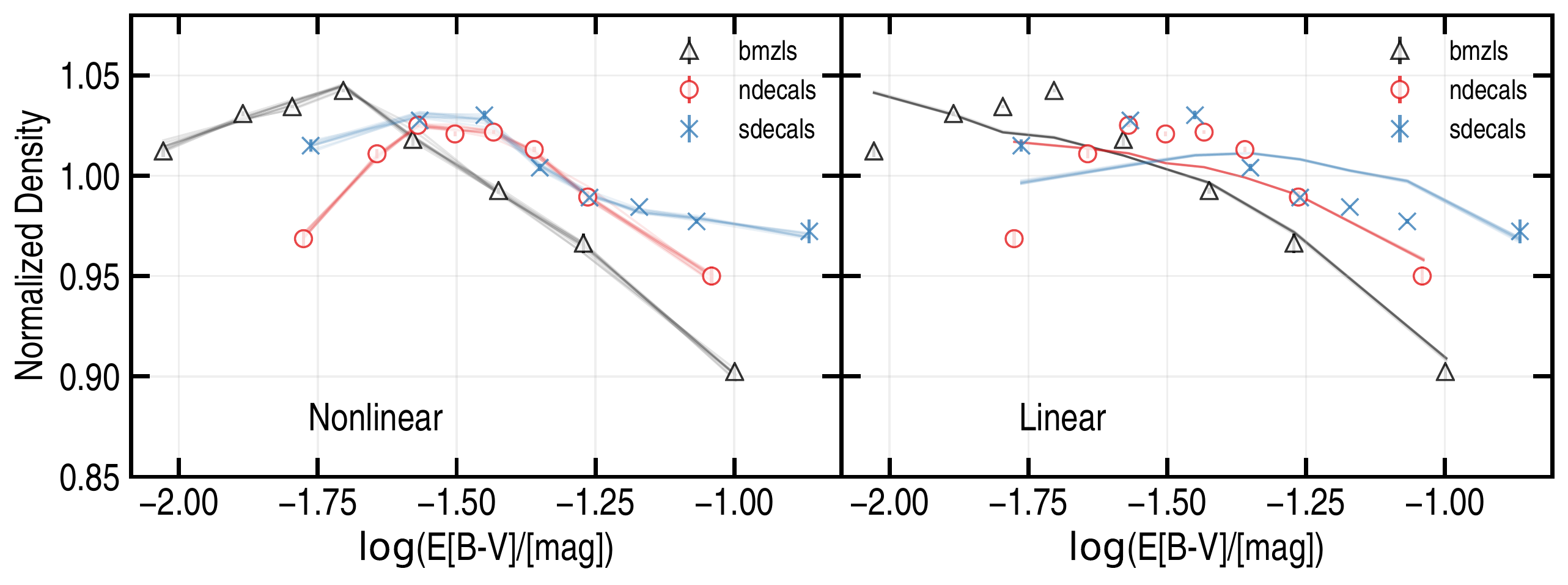}
    \caption{Normalised density of the DR9 ELGs as a function of Galactic extinction (E[B-V]) for the BASS/MzLS, DECaLS North, and DECaLS South regions. Predicted densities from the nonlinear and linear ensemble are shown in the left and right panels, respectively.}
    \label{fig:meanden_ebv}
\end{figure}

Figure~\ref{fig:meanden_ebv} shows the relative performance of the linear model and the neural network-based model; while the neural network model can learn the correlation between observed ELG number density and Galactic extinction well, the linear regression model struggles to perform as accurately.

\section{Method}
\label{sec:method}

In this section, we describe how we generate mocks with the same baseline cosmology but differing galaxy window functions to answer the questions posed in the Introduction.

\subsection{Mock Generation}
\label{sec:mocks}

\begin{table}
    \centering
    \begin{tabular}{c|c}
        \hline
        Power spectrum parameter & Value \\
        \hline 
         $h$ &  $0.6774$ \\
         $\Omega_b$ & $0.0486$ \\
         $\Omega_m$ & $0.3075$ \\
         $\Omega_K$ & $0.0$ \\
         $\Omega_R$ & $0.0015$ \\
         $A_{s} \times 10^9$ & $2.097$ \\
         $m_{\nu}$ & [$0$, $0$, $0.6$] eV \\
         $\tau$ & $0.06$ \\
         $n_s$ & $0.965$ \\
         $w_0$ & $-1$ \\
         $w_a$ & $0$ \\
         $T_{\rm CMB}$ & $2.7255$ K \\
         $N_{\rm eff}$ & $3.046$ \\
         Galaxy linear bias, $b_0$ & $1.4$ \\
         Magnification bias, $\alpha$ & $2.225$
    \end{tabular}
    \caption{Power spectrum parameters used for mock generation.}
    \label{tab:cosmo_params}
\end{table}

We run both the linear and neural network-based regressors to obtain $1000$ realizations of the galaxy window function that essentially samples the space of the galaxy window functions. We run the regressions on the observed ELG number density map from our in-preparation companion paper \citet{Karim2023}, and the feature maps are based on $11$ foreground systematics maps described in Section~\ref{sec:selfunc}. Once we obtain realizations of the window functions based on both the linear regression and neural network-based approaches, we use them to generate independent contaminated mocks of the observed galaxy overdensity field and study the impact of window variation and choices of window modelling strategies. 

At first, we generate $1000$ mocks of galaxy overdensity field, $\delta_g$, and the CMB convergence field, $\delta_{\kappa}$, with \textsc{NSIDE} = $1024$ using \textsc{Skylens} and \textsc{healpix}. \textsc{Skylens}\footnote{https://github.com/sukhdeep2/Skylens\_public} \citep{Singh21} is a theory code that calculates both the expected angular power spectra, $C_{\ell}$, and the pseudo-angular power spectra, $D_{\ell}$, given cosmology, tracer types, matter power spectra at different redshifts, redshift distribution of tracers, and the tracer window functions. It is comparable to CCL\footnote{https://github.com/LSSTDESC/CCL} but also has the functionality of calculating the coupling matrix based on the survey window, thus forward modelling what the expected pseudo angular power spectra, $D_{\ell}$ ought to be. We use \textsc{camb} \citep{camb} to calculate the matter power spectra.  

We use the same cosmology with parameters specified in Table~\ref{tab:cosmo_params} to generate the mocks. Using the same cosmology allows us to isolate the effects of only the galaxy window functions. We also use the same redshift distribution and magnification bias to generate the same underlying galaxy auto-power spectrum using \textsc{skylens}. The redshift distribution used in this analysis is estimated by calibrating the ELG colours of the DESI SV3 catalogue with the DESI Legacy Imaging Surveys DR9 photometric catalogue. A detailed discussion on the redshift distribution and magnification bias can be found in \citep{Karim2023}.

With \textsc{skylens} we obtain the three angular power spectra arrays, the CMB convergence auto-, the galaxy auto- and the CMB convergence-galaxy cross-power spectra ($C^{\kappa \kappa}_{\ell}$, $C^{gg}_{\ell}$ and $C^{\kappa g}_{\ell}$ respectively). We then pass these power spectra to the function \textsc{synfast} in \textsc{healpix} that generates correlated Gaussian realization maps of $\delta_{\kappa}$ and $\delta_g$. We generate these maps with \textsc{NSIDE = $1024$} and \textsc{pol = False} because the maps of interest are scalar fields. Note that since the scope of this paper is focused only on understanding the impact of window function variance on the power spectra and not higher-order statistics, Gaussian mocks are sufficient. We run this procedure $1000$ times to get $1000$ pairs of realizations. 

We also generate $1000$ pairs of galaxy and CMB convergence noise maps. We estimate the galaxy noise power spectrum by calculating the shot noise based on observed ELG density in the DESI Legacy Imaging Surveys DR9 dataset, and the CMB convergence noise provided by \emph{Planck}. These power spectra are then used with \textsc{healpix} to get realizations of the galaxy and CMB convergence noise maps, $\delta^{N}_{g}$ and $\delta^{N}_{\kappa}$ respectively.

For the convergence field, we add the realizations $\delta^{N}_{\kappa}$ to $\delta_{\kappa}$, apply the rotated Planck CMB lensing mask, and then finally apodize the noise-contaminated convergence field with an isotropic beam apodization given by: 

\begin{align}
\label{eq:apodize}
    b_{\ell, \cos} =
    \begin{cases}
        1 & \text{if } \ell \leq \ell_{\rm cut, min}\\
        \cos \left(\frac{\pi}{2} \frac{\ell - \ell_{\rm cut, min}}{\ell_{\rm cut, max} - \ell_{\rm cut, min}} \right) & \text{if } \ell_{\rm cut, min} < \ell_{\rm cut, max} \\
        0 & \text{if } \ell \geq \ell_{\rm cut, max}
    \end{cases}
\end{align}

We rotate the original Planck mask to the Equatorial coordinate basis before using it since Planck provides the original mask in the Galactic coordinate basis. We then apodize the noise-added and masked convergence map to reduce the boundary effect. This effectively downweights the high pseudo-$C_{\ell}$ modes. We specifically choose Equation~\ref{eq:apodize} because it is a compact function that smoothly reduces the power of the map to 0 at higher multipoles. More discussion about this apodization function can be found in Equation $83$ of \citet{Singh21}.  

For the contaminated galaxy overdensity field, $\delta_{g, C}$, we prepare the six variations (Equations~\ref{eq3:expA} -- \ref{eq3:expF}), encapsulating the two galaxy overdensity estimator definitions, and the fixed window, the multiplicative and the additive biases. Each of these six variations is prepared twice, once with the linear regression-based windows and another time with the neural network-based windows. 

Note that Equations~\ref{eq3:expA} -- \ref{eq3:expF}) refer to both the true galaxy window function, $W_{g,t}$ and the estimated galaxy window function, $W_{g,e}$. Since the true window function is never known from the first principles, we treat the mean of the $1000$ sampled galaxy window functions as the true window function, $W_{g,t}$. This is because, in expectation, the estimated window functions ought to approach the true window function. Thus, we obtain two true window functions, one for linear regression and another for the neural network-based approaches. Consequently, the $1000$ sampled realizations of the galaxy window functions are treated as the estimated window function, $W_{g,e}$. Hence, we can use our simulation-based approach to properly quantify the impact of galaxy window function multiplicative and additive biases on cosmological parameter estimation. 

Thus, in total, we generate $1000$ noise-added CMB convergence maps, and $2\times 6 \times 1000$ noise-added and window-convolved galaxy over density maps. We then use the \textsc{anafast} function from \textsc{healpix} to calculate the galaxy auto- and galaxy-convergence cross-power spectra with \textsc{lmax = 1024}. We finally subtract the window noise from the $2\times 6 \times 1000$ galaxy auto-power spectra using Equations~\ref{eq:implicit_noise} and \ref{eq:explicit_noise}. The final window-noise subtracted galaxy auto-power spectra and the galaxy-convergence cross-power spectra serve as the data vector for our analysis. 

\subsection{Validation of Mocks}
\label{sec:validation}

\begin{figure}
    \centering
    \includegraphics[width=\columnwidth]{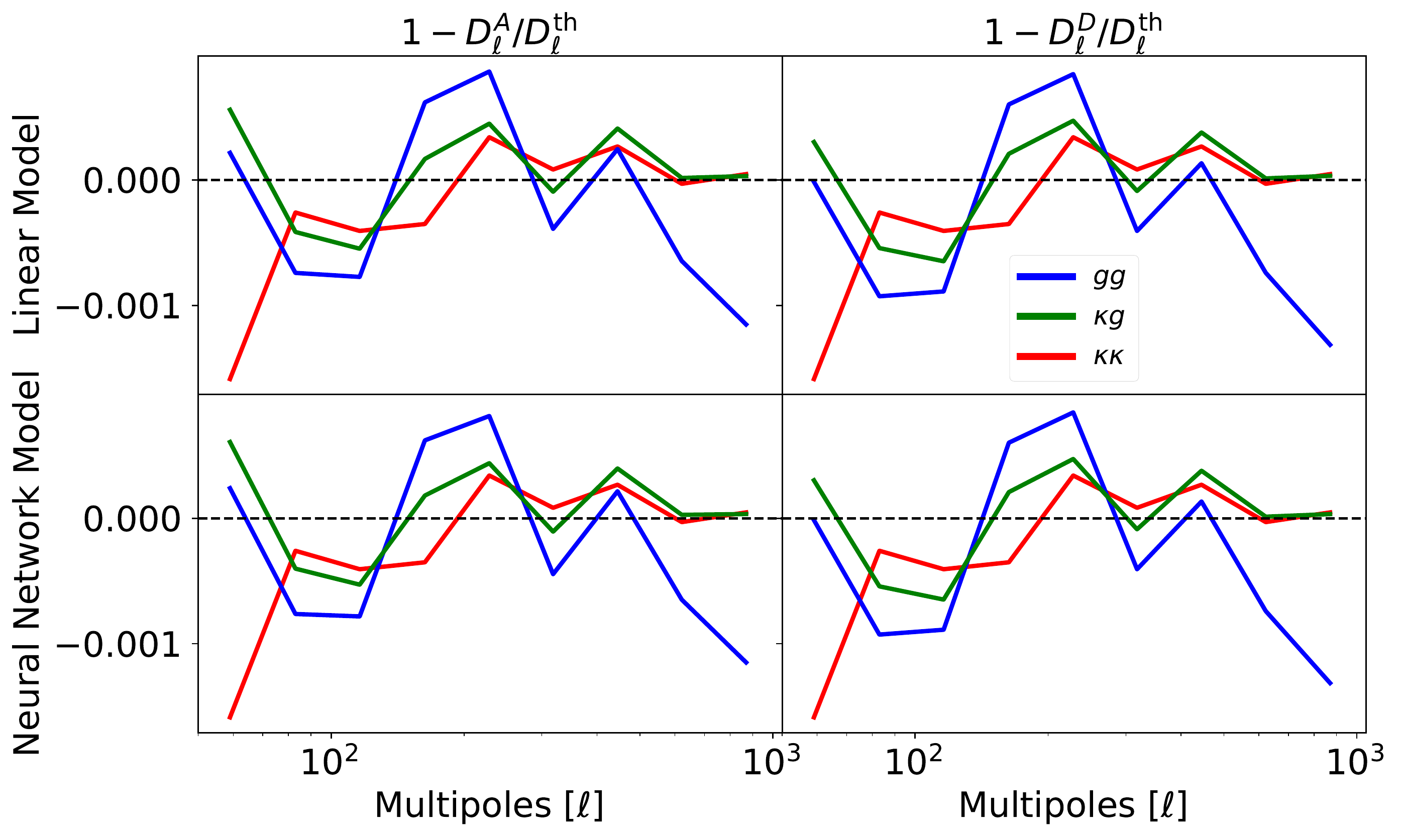}
    \caption{Comparison of the mean power spectra of the three tracers of the fixed window mocks generated using Equations~\ref{eq3:expA} and \ref{eq3:expD} with theoretical expectations. All three tracers have a sub-per cent level agreement with the theory, showing that the mock generation process is valid.} 
    \label{fig:ADcomparison}
\end{figure}

An important aspect of mock generation is to ensure that the realizations we obtain are accurate. We do this by comparing power spectra and covariance matrix measurements from the simulations with analytical predictions. Our overall goal was to achieve sub-per cent level accuracy in the mocks. We perform two different tests to show the consistency of our mocks with expectations. 

First, mocks A and D (Equations~\ref{eq3:expA} and \ref{eq3:expD} respectively) keep their respective window functions constant, and the variance only comes from the Gaussian realizations of the $\delta_g$ overdensity field. Since the underlying true power spectra and the window is known, the expected pseudo angular power spectra, $D_{\ell}$, can be calculated using Equation~\ref{eq:pcl}. Hence, we calculate the expected $D_{\ell}$ for both mocks A and D using \textsc{SkyLens} and compare the results against the mean power spectra from these mocks. As seen in Figure~\ref{fig:ADcomparison}, the disagreement between theory and the mocks is at the sub-per cent level in the multipole range of our analysis.

\begin{figure}
    \centering
    \includegraphics[width=\columnwidth]{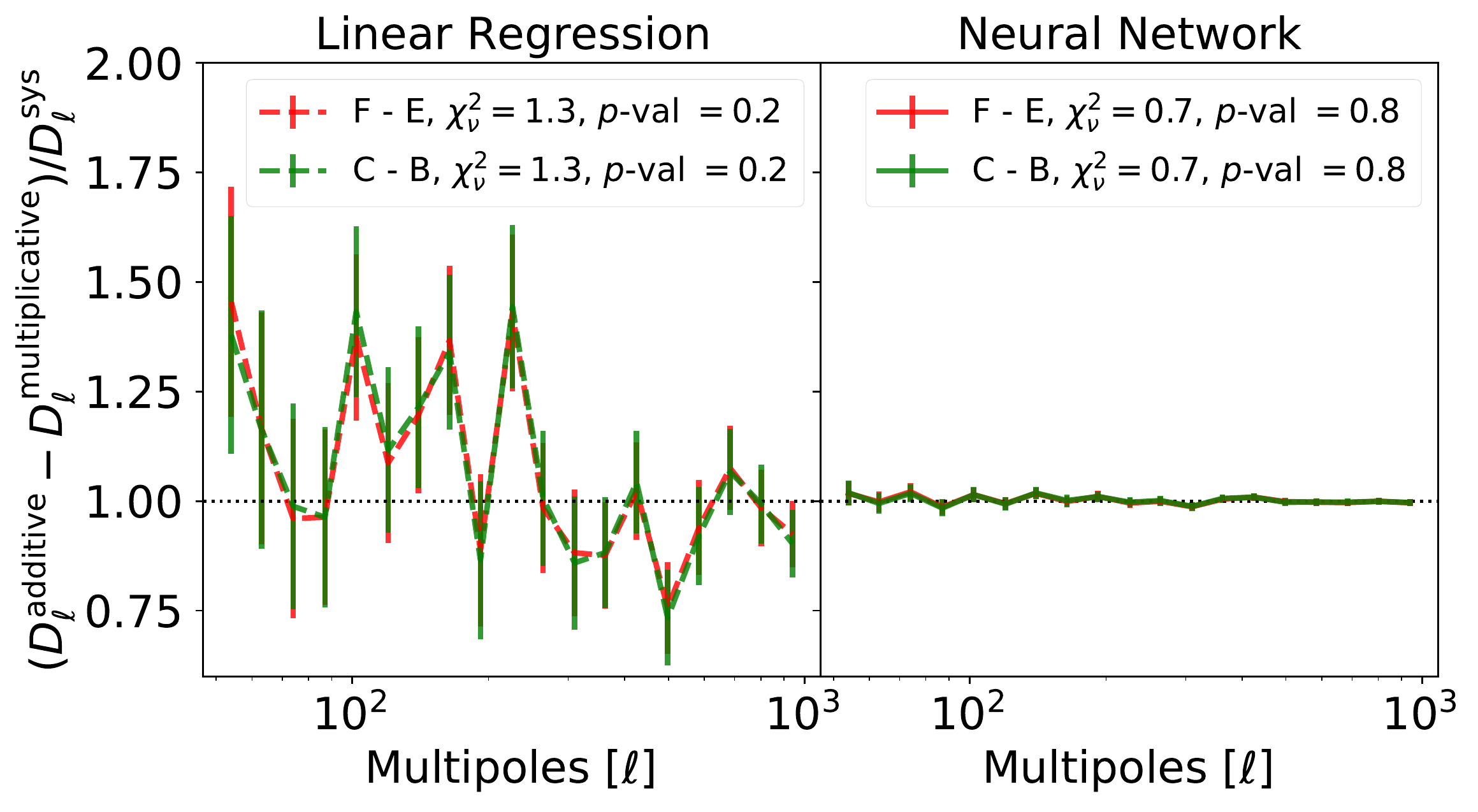}
    \caption{Comparison of the mean power spectra of the simulation-derived additive and multiplicative bias models versus theoretical expectations, based on Appendix~\ref{app:validateCF}. The plots show that both the linear regression and the neural network models agree with expectation, and validate the generative procedure of the mocks with multiplicative and additive biases embedded in them. The errorbars in the linear regression model are larger because the difference between the additive and the multiplicative components are much smaller compared to that of neural network ($\sim 100 \times$ smaller), making the ratio of the difference and the $D_{\ell}^{\rm sys}$ noisy.}
    \label{fig:CFcomparison}
\end{figure}

Once the mocks with fixed windows are validated, we then come up with a technique to validate mocks B, C, E and F that have multiplicative and additive biases added to them. In Appendix~\ref{app:validateCF} we derive analytical expressions relating mocks with only multiplicative biases, and both multiplicative and additive biases.  We compare the expected excess contribution of the additive bias with our mocks in Figure~\ref{fig:CFcomparison}. As we see here too, the generated mocks and theoretical modelling of the additive bias agree. The errorbars in the linear regression model are larger because the difference between the additive and the multiplicative components are much smaller compared to that of neural network ($\sim 100 \times$ smaller), making the ratio of the difference and the $D_{\ell}^{\rm sys}$ noisy. Thus, our generated mocks are accurate for the relevant analyses we discuss in this paper.

\section{Results}
\label{sec:result}


As stated in Section~\ref{sec3:intro}, the goal of the paper is to provide answers to three questions -- how the linear regression versus neural network regression models compare to each other when it comes to estimating the window function, whether modelling the window explicitly matters and how the bias (if any) and the variance of the estimated window function affect our final cosmological parameter inference. In this section, we answer these key questions with the simulation data.

\subsection{Linear Regression versus Neural Network Models}

\begin{figure*}
    \centering
    \includegraphics[width=0.9\textwidth]{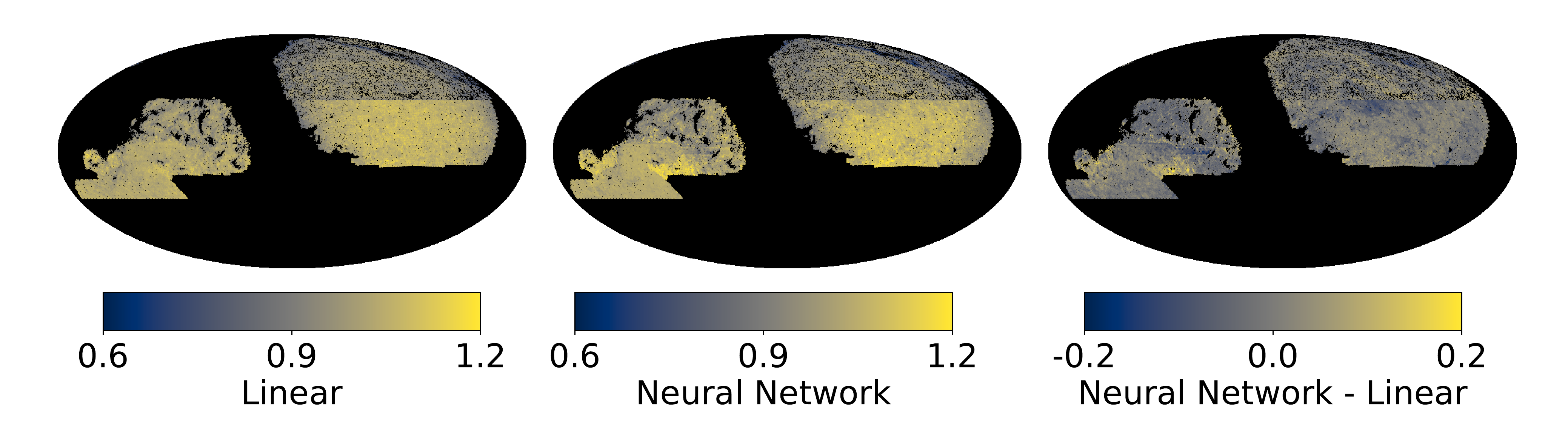}
    \caption{Mean of Window Functions based on Linear Regression and Neural Network regression methods, and their differences. The Neural Network model can learn the survey geometry of complex regions such as the DES footprint inside the Southern DECaLS footprint, which the linear regression model fails to do so.}
    \label{fig:map_compare}
\end{figure*}

The main trade-off between linear regression and neural network models is their complexities in interpretability. The linear regression model offers a straightforward insight into understanding how different underlying feature maps contribute to the galaxy window function, which is not as apparent in a neural network model. Thus, using a neural network regression model only makes sense if it can "learn" the complexities of the data that the linear regression model is unable to. 

One of the main complexities and challenges of the Legacy Surveys and DESI is that they are composed of three different distinct surveys, with the Southern DECaLS also containing the DES footprint, which has a characteristically different survey depth than the rest of the Southern DECaLS footprint. Thus, our window function should be able to model these non-cosmological variations to provide an unbiased estimate of the galaxy function. 

In Figure~\ref{fig:map_compare}, we show a comparison of the mean window functions from the $1000$ realizations of the linear regression and neural network-based approaches, and the differences of the means. We see that both of these methods give more weight to the DECaLS footprint, compared to the BASS/MzLS footprint. This makes sense since the BASS/MzLS footprint has a shallower survey depth on average. However, when we look at the DECaLS region, especially the southern DECaLS footprint, we notice a stark difference. While the linear regression model largely treats the lower half of the region as uniform, the neural network shows the DES footprint embedded within DECaLS. This is especially apparent when we look at the rightmost plot (the difference between the mean maps), where the DES footprint is visible. We also see that while the linear regression is more aggressive towards the Galactic Plane, the neural network model takes a more conservative approach in giving less weight to those same regions (closer to the borders of the surveys). This makes sense because as one observes closer to the Galactic Plane, the dust extinction on average goes up. Thus, the neural network map provides a more robust approach to modelling the complexity of the non-cosmological systematics that impact the ELG number density. 

The accuracy of the neural network model comes with a higher variance due to the bias-variance tradeoff.

There are two important consequences of this result -- both the measured power spectra and the covariance matrix are affected. 

\begin{figure}
    \centering
    \includegraphics[width=
    \columnwidth]{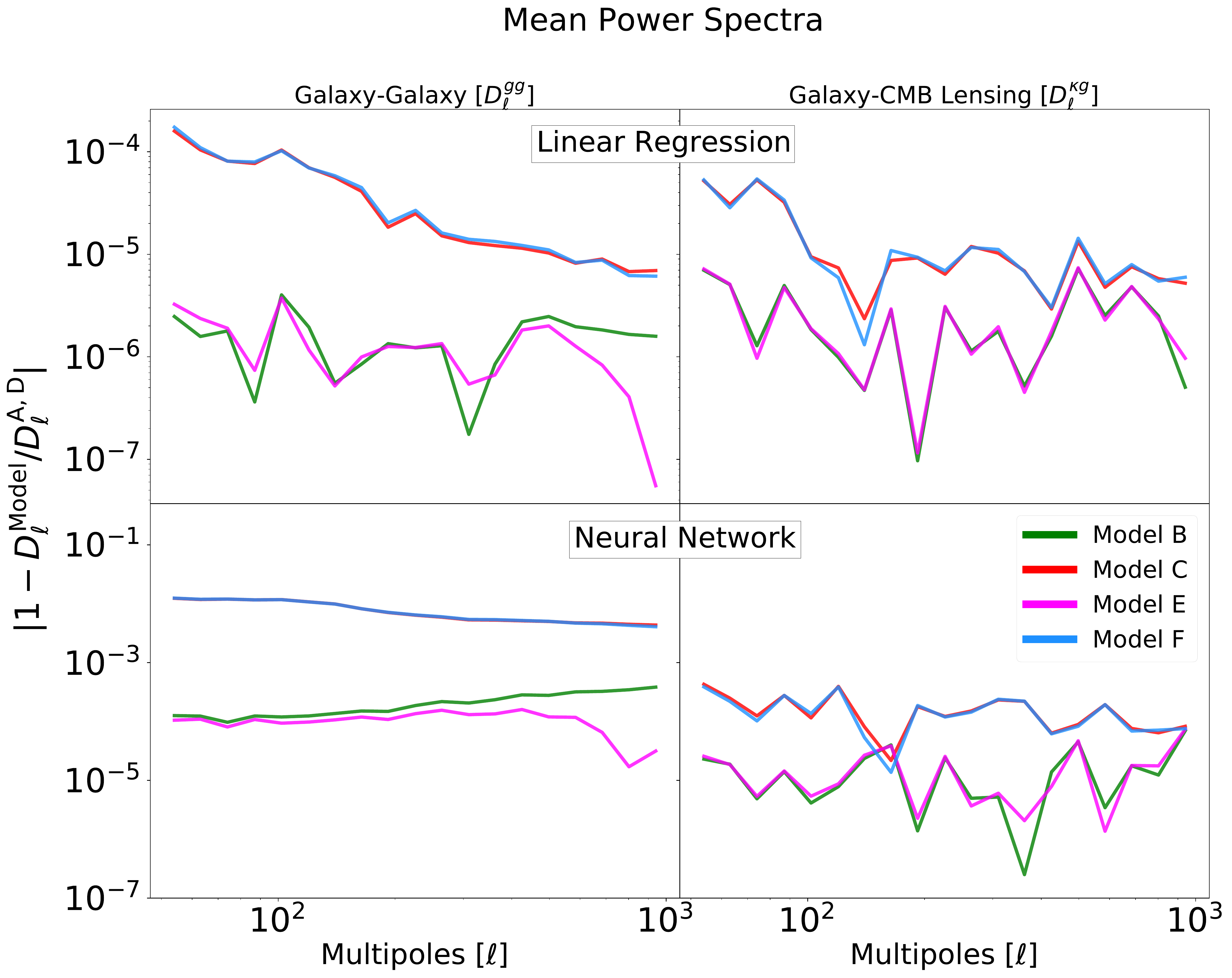}
    \caption{Mean power spectra of mocks with multiplicative and additive biases with respect to the mean of mocks with fixed windows. The ratios of Model B and C are taken with respect to Model A, and the ratios of Model E and F are taken with respect to Model D. While the additive bias in the linear regression models is sub-per cent level in galaxy auto-power spectra, the additive bias is a per cent-level effect up to $\ell \sim 350$ in the neural network based models.}
    \label{fig:mean_spectra}
\end{figure}

Figure~\ref{fig:mean_spectra} shows the comparison of mean power spectra (both galaxy-galaxy and galaxy-CMB lensing) based on both linear regression and neural network-based models. The top panel shows that for both auto- and cross-power spectra, the amount of multiplicative and additive biases are well below the per cent level at all scales of interest. This would cause one to erroneously believe that full modelling of the window function posterior is not necessary. However, the bottom panel shows a different picture. We see that while biases in the cross-power spectra are also well below the per cent level, the auto-power spectra do show a noticeable impact. Specifically in the range of $50 \lesssim \ell \lesssim 350$, the additive bias is over $1\%$. This is an important result because as modern cosmological surveys such as DESI strive for $\mathcal{O} (0.1\%)$ level precision, a per cent-level bias in the power spectra can lead to the wrong cosmological parameter inference. This is especially important for analyses that use the amplitude information to constrain cosmology, e.g. constraining $\sigma_8$ by taking a ratio of the auto- and the cross-power spectra. A similar conclusion can be drawn from Figure~\ref{fig:CFcomparison} which shows that the additive component based on the linear regression estimator is negligible, compared to the neural network estimator; this is due to the fact that the errorbars in seen in the linear regression model in Figure~\ref{fig:CFcomparison} are much larger, owing to the fact that the difference between the additive and the multiplicative bias is two orders of magnitude smaller than the difference between the additive and the multiplicative bias measured by the neural network estimator.

\begin{figure*}
    \centering
    \includegraphics[width=\textwidth]{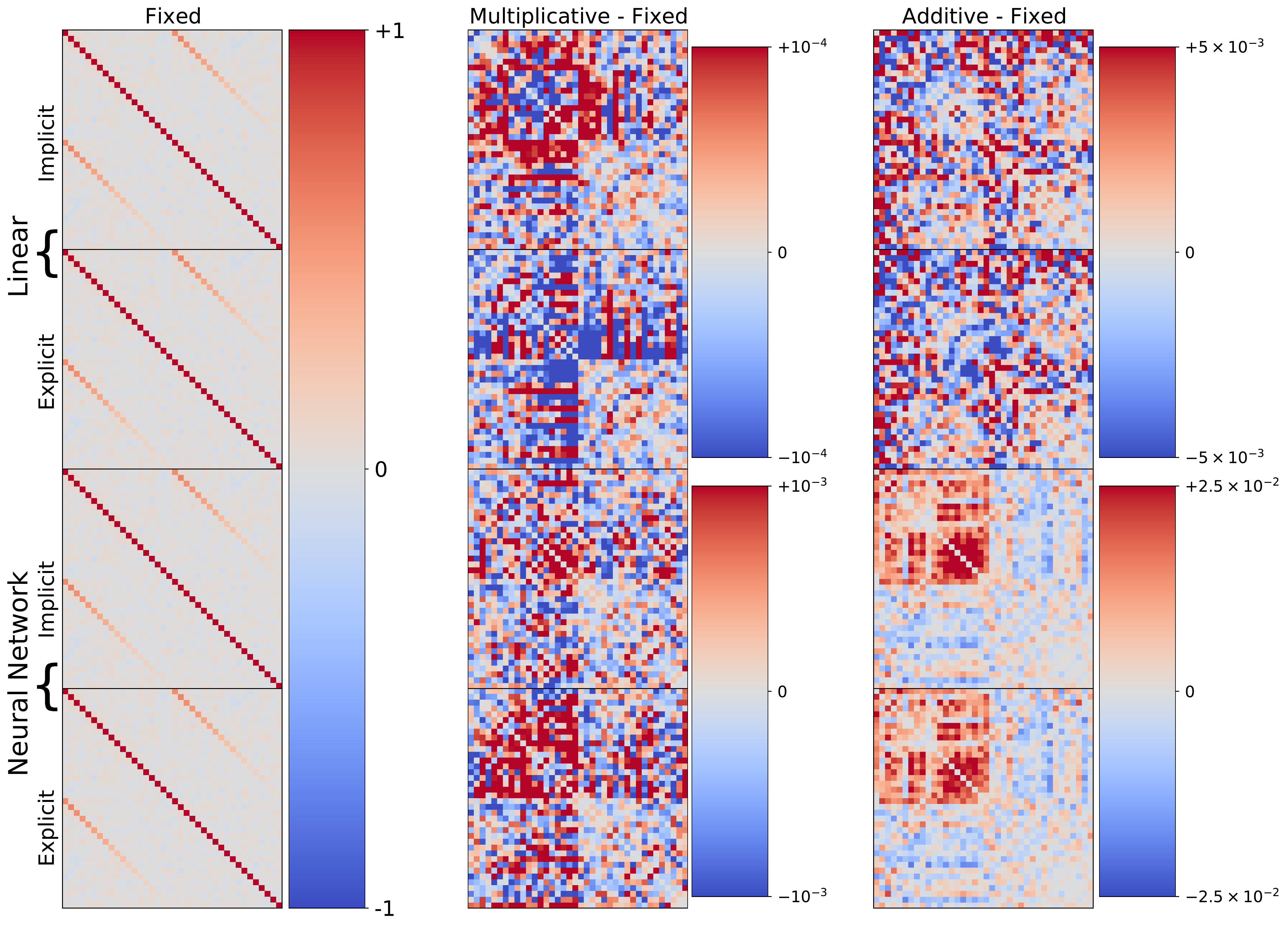}
    \caption{Correlation matrix of all the models studied in this paper. The first two rows show results from the linear regression model and the last two show results from the neural network model. The first and the third row use the implicit estimator based on Equation~\ref{eq:implicit}, and the second and the fourth row use the explicit estimator based on Equation~\ref{eq:explicit}. The first column shows the correlation matrix of fixed windows (Mocks A and D). The second column shows the difference between mocks with multiplicative bias (Mocks B and E) and the fixed window mocks, while the third column shows the difference between mocks with multiplicative and additive bases (Mocks C and F) and the fixed window mocks. The multiplicative biases are really small, while the additive biases in the neural network model are an order of magnitude larger than the additive biases seen in linear regression. Moreover, the additive bias correlation matrices show high mode-mode coupling in the galaxy auto-correlation part of the correlation matrix (top-left). This indicates that using an analytical Gaussian covariance matrix (similar to the first column) will be optimistic and measure a signal-to-noise than reality.}
    \label{fig:corrmatrix}
\end{figure*}

Additionally, the tight posterior explored by the linear regression model can artificially reduce the mode-mode coupling due to the window function. Figure~\ref{fig:corrmatrix} shows the correlation matrices based on the $1000$ realizations of all the models based on both linear regression and neural networks. The first column is the fixed window models, which as expected, show highly diagonal matrices, as expected from analytical Gaussian covariance theory. The second and third columns show the difference between the correlation matrices estimated from the biased models and the fixed window models.  The second column (models with multiplicative biases) shows a large agreement with the fixed model, indicating that multiplicative biases are not a dominant source of noise. However, the third column (models with additive biases) shows the starkest contrast between the linear regression-based and neural network-based models. The correlation between the modes in the neural network models is about $\sim 100$ times bigger than the linear regression models. We further observe a high correlation in the galaxy auto-power spectra between modes at all scales in the neural network models, which is absent in the linear regression models. 

\begin{figure}
    \centering
    \includegraphics[width=
    \columnwidth]{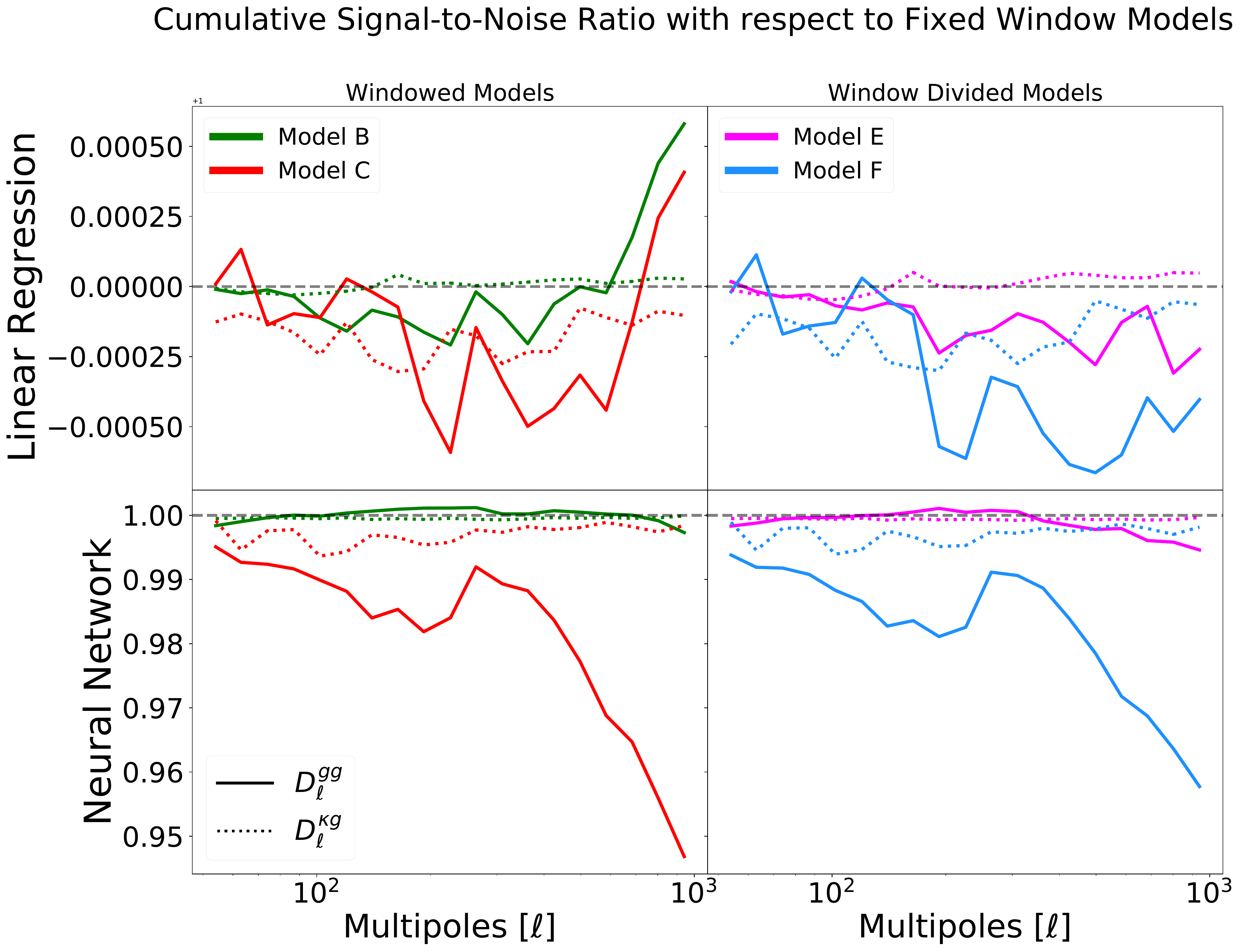}
    \caption{Signal-to-noise ratio of the mean of mocks with multiplicative and additive biases with respect to the mean of mocks with fixed windows. The ratios of Model B and C are taken with respect to Model A, and the ratios of Model E and F are taken with respect to Model D. While the linear regression-based models do not show any noticeable variation of SNR with respect to the fixed window, the neural network based models show that the additive bias affects the signal-to-noise up to $5\%$ at high multipoles.}
    \label{fig:snr}
\end{figure}

This difference has a direct consequence in our estimation of the overall cumulative signal-to-noise (SNR). Figure~\ref{fig:snr} shows that while the cumulative SNR of linear regression models is not impacted by the biases, the cumulative SNR of the neural network models is affected by the mode-mode coupling. The cumulative SNR of the galaxy auto-power spectra are affected by almost $8\%$ around $\ell \sim 1000$. This result showcases the importance of needing to model the impact of the window function on the covariance matrix because the current approaches provide overly optimistic error bars. 

\subsection{Should we model the Window function explicitly?}

The main goal of this section is to understand whether the estimator presented in Equation~\ref{eq:explicit} is better than the one presented in Equation~\ref{eq:implicit}. One of the key arguments for Equation~\ref{eq:explicit} is that if the signal is the same, the window noise component of the estimator in Equation~\ref{eq:implicit} is larger than the window noise component of the estimator in Equation~\ref{eq:explicit} \citep{Singh21}. This implies that if an analysis is dominated by the galaxy window function (or shot noise), then Equation~\ref{eq:explicit} may be the more optimal estimator. 

\begin{figure}
    \centering
    \includegraphics[width=\columnwidth]{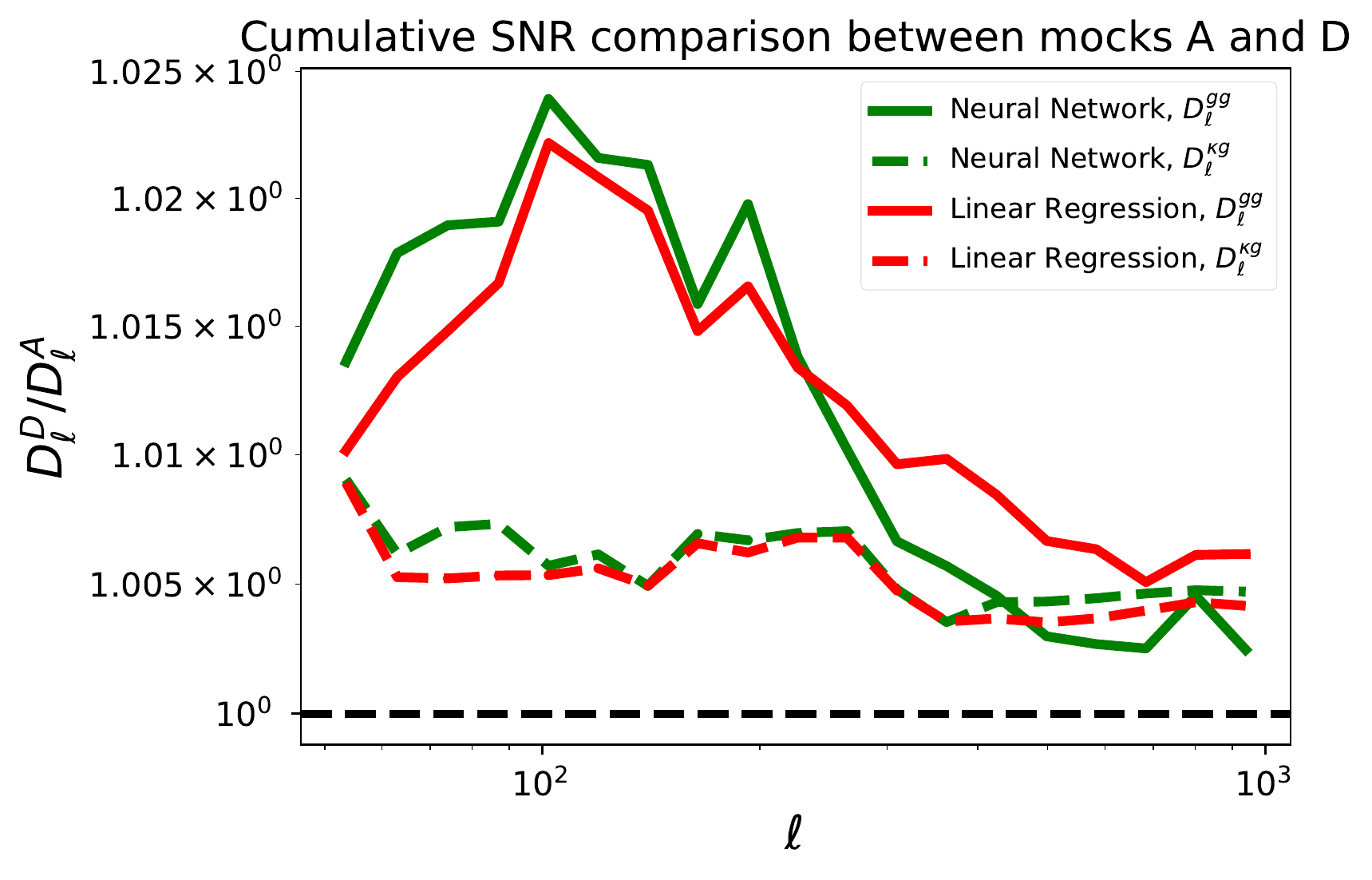}
    \caption{Cumulative signal-to-noise comparison between Mocks A and D. Mocks D has a slightly higher cumulative SNR at all scales.}
    \label{fig:snr_AD}
\end{figure}

Figure~\ref{fig:snr_AD} shows the cumulative SNR comparison between Models D (based on Equation~\ref{eq:implicit}) and Models A (based on Equations~\ref{eq:explicit}). For both linear regression and neural network-based approaches we observe that Models D have a slightly higher cumulative SNR. We pick Models A and D for this analysis because they are both fixed window models, hence we can decidedly study the effect of the signal and the noise in the presence of the same window function. 

This result makes sense because the shot noise coming from DR9 is sub-dominant at all scales of interest in our analysis. Since the signal part of the power spectrum of the estimator in Equation~\ref{eq:explicit} is affected by the variance of the window function too, in the signal-dominated regimes we expect Models D to perform slightly better. 

\begin{figure}
    \centering
    \includegraphics[width=\columnwidth]{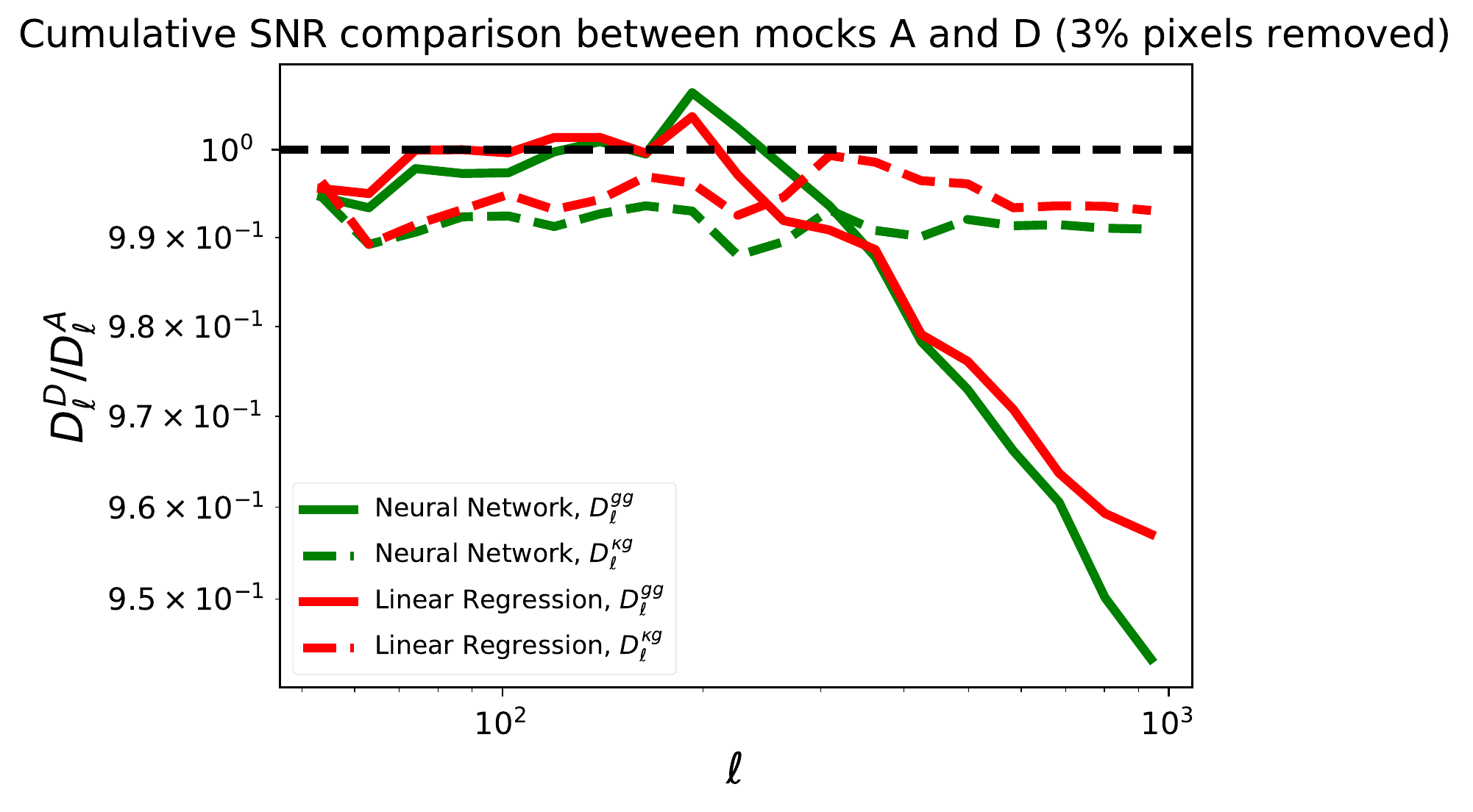}
    \caption{Cumulative signal-to-noise comparison between Mocks A and D when $3\%$ of the lowest weighted pixels are removed. This mimics the situation where pixels with low weight have to be removed when using the estimator in Equation~\ref{eq:implicit}, as opposed to the estimator in Equation~\ref{eq:explicit}. The result shows that both the galaxy auto- and galaxy-CMB lensing cross-power spectra measured using the implicit estimator have a significant reduction in the signal-to-noise compared to the same measurements made with the explicit estimator.}
    \label{fig:snr_AD_tol}
\end{figure}

However, there is another caveat to this conclusion; Model D is optimal only if the window function values are "well-behaved" in all pixels, i.e., the values are not too close to $0$. This is because the noise term in Equation~\ref{eq:implicit} is averaged over $1/W_g (x)$, and for any $x$, if $W_g (x) \to 0$, the noise term will dominate the entire estimator. Thus, it is often the case that if one uses the implicit estimator, then one must apply a further mask where any survey pixels with weights close to $0$ are removed from the analysis. For the DR9 ELGs, we found no such problematic pixels, and so Model D became the optimal estimator. But, if a hypothetical survey has many such pixels, then they need to be removed prior to measuring the power spectra; this action consequently reduces the effective survey area and can degrade the cumulative SNR. We show an example of such a situation in Figure~\ref{fig:snr_AD_tol}; we see that if we remove just $3\%$ of the lowest weighted survey footprint from the Legacy Surveys DR9, then the implicit estimator becomes severely suboptimal compared to the explicit estimator. Thus, when deciding which estimator to use, it is important to first understand the distribution of the window function weights.

\subsection{Impact of the window function modelling on cosmological parameters}

The final and perhaps the most important question of our analysis is - how do the biases in the power spectra and the differences in the covariance matrices impact cosmological parameter estimation? To answer this question, we look at three different inference problems using the fixed window model D, and the additive bias model F.  

Our ultimate goal in \citet{Karim2023} is to measure the amplitude of the power spectra, $\sigma_8$, and the galaxy linear bias, $b_0$. Assuming that the problem is Gaussian in nature, the likelihood we are maximizing is:

\begin{equation}
    \log \mathcal{L} \propto \left(M - d \right)^T \Sigma^{-1} \left(M - d \right)
\end{equation}

\noindent where, $M$ represents the proposed theoretical models that are functions of cosmological parameters, $d$ is the data vector and $\Sigma$ is the covariance matrix. As discussed in Section~\ref{sec:mocks}, we use \textsc{SkyLens}, which uses the Boltzmann solver \textsc{camb} and \textsc{HaloFits} for the linear and non-linear modelling of the matter power spectrum, to generate models represented by $M$. 

Thus with $M$ fixed, the three inference problems we study are the cases where (i) data vector, $d$, comes from mocks D and covariance, $\Sigma$, comes from mocks D, (ii) $d$ comes from mocks F and $\Sigma$ comes from mocks D, and (iii) $d$ comes from mocks D and $\Sigma$ comes from mocks F. The first inference problem tells us what is the most optimistic case, i.e., if we knew the window function exactly, how well we could constrain cosmological parameters. The second inference problem showcases what a typical survey that does not remove the additive bias and uses a covariance matrix that is devoid of window function marginalization will measure the cosmological parameters to be. And finally, the third inference shows what we ought to expect for a properly debiased data vector with a covariance matrix that takes the window function into account. We run these inferences using the Markov Chain Monte Carlo sampler, \textsc{emcee} \citep{emcee}. 

\begin{figure}
    \centering
    \includegraphics[width=\columnwidth]{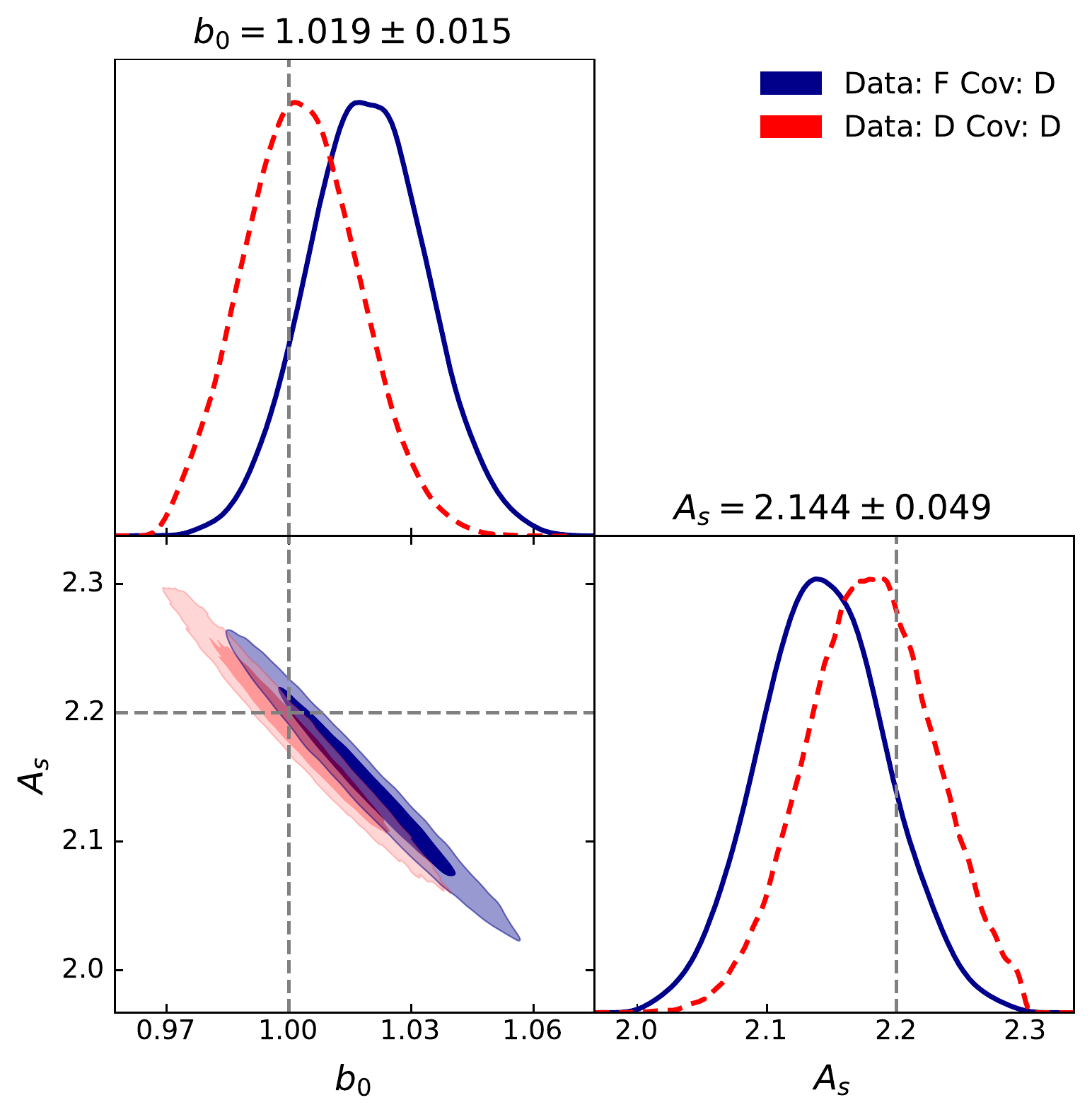}
    \caption{MCMC comparison between data vectors from mocks D and F. The dashed lines indicate the ground truth, and the red and blue contours correspond to mocks D and F respectively. In this inference problem, the covariances are the same for both cases. The figures show that the additive biases in mock F shift the mode of inferred parameters in a noticeable way.}
    \label{fig:data_DF_compare}
\end{figure}

The first comparison we conduct is the case where the same covariance matrix is used by the inference problem, but the data vectors are different. This comparison shows how much the additive bias can induce a bias in the parameter space. Figure~\ref{fig:data_DF_compare} shows that if the additive bias of the window functions is not removed, then it induces a noticeable shift in the mode of the parameters of interest. Specifically, $b_0$ and $A_s$ have a relative shift of $\sim 1.8\%$ and $\sim 2.6\%$ respectively. These numbers are significant since Stage-IV and beyond surveys will be constraining the precision to per cent level \citep{DESI16}. 

\begin{table}
\centering
\begin{tabular}{|c||c|c|} 
 \hline
 & D & F \\ [0.5ex] 
 \hline\hline
 $\sigma_{b_0^2}$ & $2.10 \times 10^{-4}$ & $2.27 \times 10^{-4}$\\ 
 \hline
 $\sigma_{A_s^2}$ & $2.54 \times 10^{-3}$ & $2.66\times 10^{-3}$\\ 
 \hline
 $\sigma_{b_0} \sigma_{A_s}$ & $-7.18\times 10^{-4}$ & $-7.63\times 10^{-4}$\\ [1ex] 
 \hline
\end{tabular}
\caption{Table showing parameter covariance comparison between mocks D and F. Compared to mocks D, mocks F have a higher covariance.}
\label{tab:DF_comparison}
\end{table}

The next comparison we do is the case where we use the same data vector but different covariance matrices. This comparison shows us how much our precision degrades due to the presence of the additive and multiplicative biases compared to the analytic Gaussian covariance matrix case. Table~\ref{tab:DF_comparison} shows that compared to mocks D (which corresponds to analytical Gaussian covariance matrix case), mocks F have overall larger covariance values. In fact, the ratio of the error ellipses between F and D is $\sim 1.246$, showing that overall the precision also goes down in a noticeable fashion. 

These results combined showcase that modelling the multiplicative and additive biases will be an important source of systematics for ongoing and future cosmological surveys.

\section{Conclusions}
\label{sec:conc}
The goal of this paper was to understand how the estimation of the (galaxy) window function impacts overall cosmological parameter inference. We motivate this analysis by studying the effects of modelling the galaxy window function on measuring the amplitude of the matter power spectrum, $A_s$ and the galaxy linear bias, $b_0$, by cross-correlating the Legacy Surveys DR9 Emission-Line Galaxies catalogue with the \emph{Planck} CMB lensing map. Our results indicate that: 

\begin{enumerate}
    \item {\bf Linear Regression versus Neural Network Methods}: Neural Network (or other non-linear regressors) based regression methods are clearly superior compared to linear regression models when it comes to modelling the galaxy window function. While linear regression models have really high precision, they also have a high bias as they cannot model the non-linear mapping between the imaging systematics maps and the corresponding observed galaxy number density map. 
    
    On the other hand, while the neural network models have a higher variance, they are more accurate and one can explicitly model the amount of bias using a simulation-based approach. For our datasets, the amount of bias in the galaxy auto power spectra, $D_{\ell}^{gg}$, is $\sim 1-2\%$ up to $\ell \sim 1000$. This result also indicates that rather than simply using the best value of the window function, one needs to sample the posterior space of the window function properly to get an unbiased estimate. 
    
    \item {\bf Modeling the window function explicitly in the overdensity estimator}: The importance of modelling the window function explicitly in the galaxy overdensity estimator is not as apparent. While \citep{Singh21} analytically showed that explicit modelling should lead to a higher signal-to-noise, it is only applicable in the noise-dominated regime. As high-number density surveys like the Legacy Surveys or DESI are signal-dominated up to significantly high multipoles, modelling the window function explicitly in the overdensity estimator is not essential. However, if a high-$\ell$ signal-dominated survey has an unusual survey footprint, in that a fraction of its footprint has the window function close to $0$, then modelling the window explicitly in the overdensity estimator does provide some advantage. 
    
    \item {\bf Marginalizing the window function}: Compared to the analytical Gaussian covariance matrix, a simulation-based covariance matrix modelling shows that the additive bias in the galaxy window function induces significant mode-mode couplings in the galaxy auto-power spectra. This results in an overall cumulative signal-to-noise reduction of $\sim 10\%$ at $\ell_{\rm max} \sim 1000$. Thus, properly modelling the posterior of the window function and marginalizing over it is essential to get the proper error bars. 
    
    \item {\bf Impact of bias and marginalization on parameter inference}: The additive bias estimated in the galaxy auto-power spectra causes a shift of $\sim 1.8\%$ and $\sim 2.6\%$ in $b_0$ and $A_s$, respectively. Moreover, properly modelling the covariance matrix increases the error ellipse of these parameters by $\sim 25\%$ compared to the case where one uses analytical Gaussian covariance.\\
\end{enumerate}

One caveat to note is that in this work we are not capturing all possible sources of error when modelling the galaxy window function. Importantly, if the imaging systematic maps themselves are wrong, then that will introduce additional bias and noise. Our analysis is only restricted to showing that even in the case where we trust all the imaging systematics maps, one still needs to model the bias and variance of the galaxy window function appropriately. 

While previous studies of the impact of the window function on cosmological studies have been limited to large-scale modes \citep{Rezaie20}, this paper for the first time shows that the bias and uncertainty of the galaxy window function has a statistically significant impact at small-scales too, due to mode-mode coupling induced by the additive bias in the galaxy window function. 

All of these results combined indicate that the bias and uncertainty of the window function require a careful and proper characterization. Since future surveys like DESI, \emph{Euclid}, \emph{Roman} and the Rubin Observatory will try to achieve sub-per cent level precision in measuring cosmological parameters using galaxy clustering, the identified systematics will be one of the dominant sources of systematic uncertainties in the overall error budget. Thus, we recommend that modelling the window function should be an essential consideration for all cosmological surveys going forward. 

\section*{Acknowledgements}

TK and DJE are supported by the U.S. Department of Energy Office of Science through DE-SC0007881 and by the Simons Foundation. TK is additionally supported by the National Science Foundation Graduate Research Fellowship under Grant No. DGE - 1745303 and the Barbara Bell Dissertation Fellowship. MR is supported by the U.S. Department of Energy grants DE-SC0021165 and DE-SC0011840. SS is supported by the McWilliams Fellowship at Carnegie Mellon University. 

The authors would like to thank the referee for their valuable feedback. The authors would also like to thank Eva Mueller, Nayantara Mudur, Douglas Finkbeiner, Lado Samushia, Hee-Jong Seo, Ashley Ross, and Reza Katebi for stimulating discussions that helped prepare this work. MR additionally would like to thank Ohio State University's Center for Cosmology and AstroParticle Physics, in particular, John Beacom and Lisa Colarosa, for their hospitality and support.

The DESI Legacy Imaging Surveys consist of three individual and complementary projects: the Dark Energy Camera Legacy Survey (DECaLS), the Beijing-Arizona Sky Survey (BASS), and the Mayall z-band Legacy Survey (MzLS). DECaLS, BASS and MzLS together include data obtained, respectively, at the Blanco telescope, Cerro Tololo Inter-American Observatory, NSF’s NOIRLab; the Bok telescope, Steward Observatory, University of Arizona; and the Mayall telescope, Kitt Peak National Observatory, NOIRLab.

This research used resources of the National Energy Research Scientific Computing Center (NERSC), a U.S. Department of Energy Office of Science User Facility located at Lawrence Berkeley National Laboratory, operated under Contract No. DE-AC02-05CH11231.

This research used the Ohio Supercomputer Center's high performance computing and data storage infrastructure \citep{OhioSupercomputerCenter1987}.

This research made use of the following packages: \textsc{Numpy} \citep{numpy}, \textsc{Healpy} and \textsc{HEALPix} \citep{Zonca2019, Gorski05}, \textsc{SkyLens} \citep{Singh21}, \textsc{CAMB} \citep{camb},  \textsc{emcee} \citep{emcee} and \textsc{GetDist} \citep{getdist}.
\section*{Data Availability}

The simulation power spectra, redshift distribution, and the final chains of the MCMC-based sampling are all stored in the National Energy Research Scientific Computing Center (NERSC) facility. The first author will make the path to the data available for analysis within NERSC, or a Globus link for external users, upon request.
 
\bibliographystyle{mnras}
\bibliography{main} 


\appendix

\section{Analytic Estimation of Excess Clustering due to Additive Bias}
\label{app:validateCF}

Assume that the estimated and the true galaxy window functions are given by $W_g$ and $W_{g,t}$ respectively. We define the true window function as being the mean of the $1000$ realizations we obtain using our regression methods. This makes sense because in expectation we would expect the window function realizations to approach the true window function.

With the true window defined, we now model the mocks that have biases induced in them. In the implicit approach, the observed $\delta_g$ field with multiplicative and additive biases are then given by Equations~\ref{eq3:expE} and \ref{eq3:expF}: 

\begin{align*}
    \delta_E &= \frac{W_g}{W_{g,t}} \delta_g + \frac{\sqrt{W_g}}{W_{g,t}}\delta_g^N\\
    \delta_F &= \frac{W_g}{W_{g,t}} \left( 1 + \delta_g \right) + \frac{\sqrt{W_g}}{W_{g,t}} \delta^N_g - 1\\
    &= \left( \frac{W_g}{W_{g,t}}  -1  \right) + \delta_E 
\end{align*}

\noindent Let us define the following quantity:

\begin{equation}
 \left( \frac{W_g}{W_{g,t}}  -1  \right) \coloneqq \delta_{\rm sys}   \label{eq:delta_sys}
\end{equation}

\noindent Plugging this back into the final expression, we get: 

\begin{equation}
    \delta_F = \delta_{\rm sys} + \delta_E
\end{equation}

We can now calculate the auto-power spectra on both sides of the expression: 

\begin{align}
    &\delta_F = \delta_{\rm sys} + \delta_E \\
&\implies \left< \delta_F \delta_F'  \right> = \left< \left( \delta_{\rm sys} + \delta_E \right) \left( \delta_{\rm sys}' + \delta_E' \right) \right> \\
&\implies D^{\ell}_{F} = \left<\delta_{\rm sys} \delta_{\rm sys}' \right> + \left<\delta_E \delta_E' \right> + 2\left<\delta_{\rm sys} \delta_E' \right>  \label{eq3:delsys_delE}\\
&\implies D^{\ell}_{F} = \left<\delta_{\rm sys} \delta_{\rm sys}' \right> + D^{\ell}_E \qquad \left[ \because \delta_{\rm sys} \perp \delta_{E} \right] \\
&\implies D^{\ell}_{F} = D^{\ell}_{\rm sys} + D^{\ell}_E \qquad \left[ D^{\ell}_{\rm sys} \coloneqq \left<\delta_{\rm sys} \delta_{\rm sys}' \right> \right]
\end{align}

\noindent Therefore, measured power spectra of mocks with both multiplicative and additive biases are related to mocks with only multiplicative biases by a simple expression. We now derive explicitly what the term $D^{\ell}_{\rm sys}$ consists of using known quantities: 

\begin{align*}
    D_{\rm sys}^{\ell} &= \left<\delta_{\rm sys} \delta_{\rm sys}' \right> \\
    &= \left< \left( \frac{W_g}{W_{g,t}}  -1 \right) \left( \frac{W^{'}_g}{W_{g,t}}  -1 \right)  \right> \\
\end{align*}

Thus, using the window realizations, we can quantify the power spectra due to the excess additive bias.

Similarly, for the explicit approach, we have expressions for the observed $\delta_g$ field with multiplicative and additive biases as Equations~\ref{eq3:expB} and \ref{eq3:expC}:

\begin{align*}
    \delta_B &= W_g\delta_g + \sqrt{W_g} \delta_g^N\\
    \delta_C &= W_g \left( 1 + \delta_g \right) + W_g \delta^N_g - W_{g,t}\\
    &= \left(W_g - W_{g,t}  \right) + \delta_B 
\end{align*}

\noindent We recognize that, $\left(W_g - W_{g,t}  \right) = \delta_{\rm sys} W_{g,t}$, where $\delta_{\rm sys}$ refers to the expression defined Equation~\ref{eq:delta_sys}. Thus,

 \begin{equation}
    \delta_C = \delta_{\rm sys} W_{g,t} + \delta_B 
\end{equation}

\noindent We can now take the auto-power spectra on both sides of this expression:

 \begin{align*}
    &\delta_C = \delta_{\rm sys} W_{g,t} + \delta_B \\
    &\implies \left< \delta_C \delta_C'  \right> = \left< \left( \delta_{\rm sys}W_{g,t} + \delta_B \right) \left( \delta_{\rm sys}' W_{g,t} + \delta_B' \right) \right> \\
&\implies D^{\ell}_{C} = \left<\delta_{\rm sys} W_{g,t} \delta^{'}_{\rm sys} W_{g,t} \right> + \left<\delta_B \delta_B' \right> + 2\left<\delta_{\rm sys} W_{g,t} \delta_B' \right>  \\
&\implies D^{\ell}_{C} = \left<\delta_{\rm sys} W_{g,t} \delta_{\rm sys}' W_{g,t} \right> + D^{\ell}_B \qquad \left[ \because \delta_{\rm sys} \perp \delta_B \right] \\
&\implies D^{\ell}_{C} = D^{\ell}_{\rm sys'} + D^{\ell}_B
\end{align*}

\noindent Again, we see that the power spectra of mocks with both multiplicative and additive biases are related to mocks with only multiplicative biases by a simple expression. 

Thus, the systematics power spectra are given as:

\begin{align}
    D_{\rm sys}^{\ell} &= \left< \left( \frac{W_g}{W_{g,t}}  -1 \right) \left( \frac{W^{'}_g}{W_{g,t}}  -1 \right)  \right> \label{eq3:sys_EF} \\
    D^{\ell}_{\rm sys'} &= \left<\delta_{\rm sys} W_{g,t} \delta_{\rm sys}' W_{g,t} \right> \label{eq3:sys_BC}
\end{align}

The denominator in Figure~\ref{fig:CFcomparison} are the expressions~\ref{eq3:sys_EF} and \ref{eq3:sys_BC} respectively. The comparisons in Figure~\ref{fig:CFcomparison} show error bars because the expressions we have derived in this section are true only in expectation, but not necessarily in per realization. Thus, we take this effect into account when performing the comparison.


\bsp	
\label{lastpage}
\end{document}